\documentclass[prb,reprint,nofootinbib]{revtex4-1}
\bibpunct{[}{]}{;}{n}{}{}

\pdfoutput=1
\usepackage[caption=false]{subfig}

\usepackage{graphicx}
\usepackage{comment}
\usepackage{amsmath}
\usepackage{amssymb}
\usepackage{amsfonts}
\usepackage{bbm}
\usepackage{braket}
\usepackage{verbatim}
\usepackage[hidelinks]{hyperref}
\usepackage{ulem}
\hypersetup{
  colorlinks   = true, 
  urlcolor     = blue, 
  linkcolor    = blue, 
  citecolor   = blue 
}

\usepackage{color}
\usepackage{tikz}

\newcommand{\be}{\begin{equation}}
\newcommand{\ee}{\end{equation}}

\newcommand{\tr}[1]{\text{Tr}\big[{#1}\big]}

\begin{document}
	
	\title{Variational Schrieffer-Wolff Transformations for Quantum Many-Body Dynamics}
	
	\author{Jonathan Wurtz}
	\email[Corresponding author: ] {jwurtz@bu.edu}
	\affiliation{Department of Physics, Boston University, 590 Commonwealth Ave., Boston, MA 02215, USA}
	
	\author{Pieter W. Claeys}
	\affiliation{Department of Physics, Boston University, 590 Commonwealth Ave., Boston, MA 02215, USA}

	\author{Anatoli Polkovnikov}
	\affiliation{Department of Physics, Boston University, 590 Commonwealth Ave., Boston, MA 02215, USA}

	\begin{abstract}
	Building on recent results for adiabatic gauge potentials, we propose a variational approach for computing the generator of Schrieffer-Wolff transformations. These transformations consist of block diagonalizing a Hamiltonian through a unitary rotation, which leads to effective dynamics in a computationally tractable reduced Hilbert space. The generator of these rotations are computed variationally and thus go beyond standard perturbative methods; the error is controlled by the locality of the variational ansatz. The method is demonstrated on two models. First, in the attractive Fermi-Hubbard model with on-site disorder, we find indications of a lack of observable many-body localization in the thermodynamic limit due to the inevitable mixture of different spinon sectors. Second, in the low-energy sector of the XY spin model with a broken $U(1)$-symmetry, we analyze ground state response functions by combining the variational SW transformation with the truncated spectrum approach.
	\end{abstract}

	\maketitle

	\section{Introduction}
	
    One of the main challenges in quantum theory is computing the dynamics of involved quantum systems without having to resort to exact diagonalization or conventional perturbation theory \cite{Griffiths2005}. While exact dynamics are generally out of reach, one potential way of obtaining approximate dynamics is through a  basis rotation, simplifying the rotated Hamiltonian and the resulting dynamics in the new frame. An extreme example is going to the eigenbasis of the Hamiltonian, where the dynamics are trivial -- each state simply picks up a phase. Unfortunately, this basis is generally inaccessible due to the prohibitively large Hilbert space, and approximate methods need to be found. One such alternative method is the Schrieffer-Wolff (SW) transformation \cite{Schrieffer1966a,Bravyi2011a}. Provided there's a clear separation of energy scales within a given Hamiltonian, one finds a unitary transformation block-diagonalizing and thus decoupling the low- and high-energy subspaces of the model. The low-energy dynamics then follow from an effective SW Hamiltonian. However, the traditional way of implementing the SW transformation is perturbative and can be used only if there is a very large energy scale separation, otherwise one quickly encounters the problem of small denominators. Mapping a static Hamiltonian to a Floquet problem in the rotating frame and applying the van Vleck high frequency expansion, which was shown to be equivalent to the SW transformation~ \cite{Bukov2016a}, one deals with an asymptotic series that also becomes uncontrollable in the absence of such a large energy scale separation~\cite{Abanin2015, Mori2016}. Similar principles \cite{Vogl2019} underlie the Wegner flow, where a flow equation is constructed band-diagonalizing the Hamiltonian through the systematic suppression of off-diagonal matrix elements associated with smaller and smaller energy differences \cite{FranzWegner1990a}.

These diagonalization methods can be reinterpreted in the context of adiabatic gauge potentials (AGPs) \cite{Kolodrubetz2017}, which are infinitesimal generators of a unitary transformation diagonalizing a given Hamiltonian. Recent works have allowed for controllable variational approximations to the AGP, which lead to unitary transformations  partially diagonalizing the Hamiltonian \cite{Sels2016b,Kolodrubetz2017,Claeys2019a}. The variational AGP is guaranteed to converge to the exact one if the number of variational parameters becomes sufficiently large. In practice, the convergence properties depend on the details of the Hamiltonian, the choice of the variational manifold, the particular energy sector one is interested in, and so on. But even with these limitations, the variational SW transformation has a clear advantage over the perturbative expansions, which generally have a zero radius of convergence. This advantage stems from the fact that the generator of the rotation can be stably computed at any value of the couplings. In this work, we show how Hamiltonians rotated using variational AGPs allow for accurate simulations of dynamics at a fraction of the cost of exact methods. This methodology then allows a description of low-energy quenches and other effective dynamics of interacting quantum systems in non-perturbative regimes.

More specifically, we start from an initial Hamiltonian, which is easily diagonalizable and naturally splits into degenerate or nearly-degenerate blocks. Then we introduce a coupling which breaks this block structure and lifts (near-)degeneracies. Using a variationally-approximated AGP we perform a unitary rotation which approximately restores the block-diagonal structure of the Hamiltonian and then project it to a subspace containing the relevant degrees of freedom (for example, low-energy states). This projection leads to an effective Hamiltonian with drastically reduced Hilbert space, making exact dynamics accessible.

This method is illustrated on two classes of systems. First, we consider a disordered strongly-attractive Fermi-Hubbard model which, using the lowest order SW transformation, can be mapped to the disordered Heisenberg Hamiltonian~\cite{Blundell2001}, where empty and doubly-occupied states form effective spin degrees of freedom. If the disorder is sufficiently large this model exhibits many-body localization, which is manifested in absence or near absence of thermalization \cite{Abanin2019}. Using the variational approach we go beyond this perturbative construction and obtain a more accurate effective Hamiltonian, which contains a mixture of singly-occupied sites. In turn this mixture leads to enhanced transport in the system and restores thermalization in the system.

Next, we apply this method to a non-integrable Ising model with transverse and longitudinal field and calculate response functions above the ground state. We apply a similar strategy to rotate the Hamiltonian, effectively eliminating the integrability-breaking longitudinal field and replacing it with longer range spin-spin interactions. Then, a truncated spectrum approach is used to perform the necessary projection of the transformed Hamiltonian to the low-energy subspace. In this way we can enter a deeply non-integrable and non-perturbative regime, where standard SW and truncated spectrum approaches are not applicable.

	\section{Methodology}
	
	\begin{figure}
        \centering
        \includegraphics[width=\linewidth]{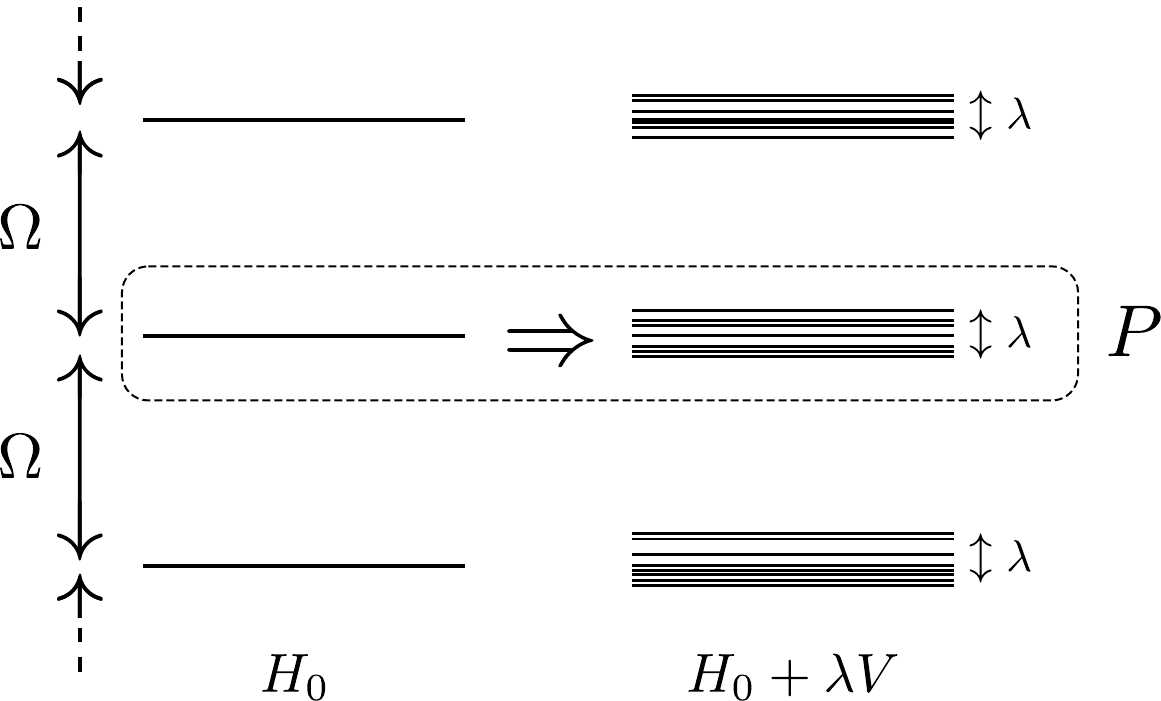}
        \caption{A Hamiltonian $H_0$ with equidistant highly-degenerate energy levels separated by $\Omega$. An extra term $\lambda V$ breaks this degeneracy and induces level splitting on the order of $\lambda$, but dynamics within a given subspace $P$ may still be well-described via a SW rotation.}
        \label{fig:SW_schematic}
    \end{figure}

	The goal of the proposed approach corresponds to that of the Schrieffer-Wolff (SW) transformation \cite{Bravyi2011a}: given a Hamiltonian acting on different subspaces, an effective Hamiltonian is found acting only on a single subspace, integrating out the degrees of freedom from other subspaces. These are generally taken to be low- and high-energy subspaces (as in Fig.~\ref{fig:SW_schematic}), leading to an effective low-energy Hamiltonian. Given an unperturbed Hamiltonian $H_0$ with well-separated energy subspaces (in the Figure represented by highly-degenerate levels separated by an energy scale $\Omega$), a term added to this Hamiltonian as
	\begin{equation}
	    H = H_0 + \lambda V
	\end{equation}
will break this degeneracy and lead to level-splitting in the spectrum of the total Hamiltonian. Here, we assume the strength of the perturbation $\lambda$ to be small enough such that the mixing between different degenerate sectors of $H_0$ is not very strong (loosely speaking, $\lambda < \Omega$). Starting from a state within a subspace $P$, the dynamics of the full model will mainly be governed by the Hamiltonian acting within this subspace, with states within the complement $Q$ of this subspace only leading to small high-frequency deviations. The goal of the SW transformation is to find an effective Hamiltonian acting only within $P$ that is able to describe these dynamics. Conventionally the SW transformation splits into three steps.

First, some projective subspace $P$ is identified in which the Hamiltonian $H_0$ is block-diagonal. This could be a specific energy sector(s) (as in Fig.~\ref{fig:SW_schematic}), or alternatively some symmetry sector(s) of $H_0$. Second, one finds a unitary rotation $U$ to a new basis ``$\sim$" such that the Hamiltonian transformed by this unitary $\tilde H = U^{\dagger}HU$ is block-diagonal in $P$ or, equivalently, the original Hamiltonian is block-diagonal in the transformed basis $\tilde{P}$  (see also Fig.~\ref{fig:block_diagonal_diagram}),
	\begin{align}\label{eq:Htilde_definition}
	\tilde H = U^\dagger HU\quad&\leftrightarrow\quad &\mathcal P \tilde H \mathcal Q + \mathcal Q\tilde H\mathcal P=0,\\
	\tilde{\mathcal P} = U\mathcal PU^\dagger\quad&\leftrightarrow &\mathcal{\tilde P}H\mathcal{\tilde Q} + \mathcal{\tilde Q} H \mathcal{\tilde P}=0, 
	\end{align}
    where $\mathcal P,\mathcal Q$ stand for the projectors to the subspaces $P$ and $Q$, respectively.    Third, an effective Hamiltonian is constructed as a projection of $\tilde H$ into the block $P$
	\begin{equation}
	    \tilde H_\text{eff} = \mathcal P \tilde H\mathcal P.
	\label{eq:HeffProj}
	\end{equation}
This Hamiltonian effectively projects out the degrees of freedom outside of $P$ such that the dynamics of wave functions with overlap predominantly in subspace $\tilde{P}$ can then be described in terms of this effective Hamiltonian. As such, this new Hamiltonian $\tilde H_\text{eff}$ has the clear advantage that it acts on a reduced Hilbert space, which can be substantially smaller than that of the original Hamiltonian.

	\begin{figure}
		\includegraphics[width=0.65\linewidth]{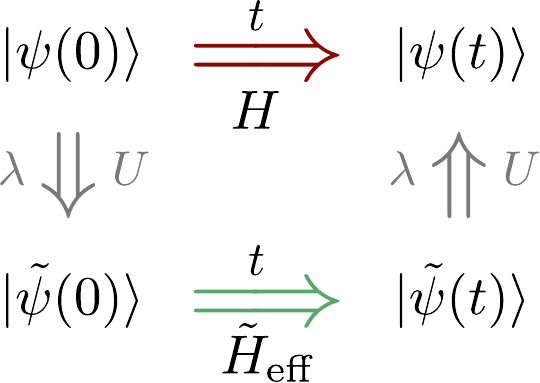}
		\caption{Given an initial wave function $\ket{\psi(0)}$, we wish to find time evolution $\ket{\psi(t)}$ with respect to Hamiltonian $H$ (top). However, this may be intractable. Instead, after a unitary transformation w.r.t. $U$ of both the wave function and the Hamiltonian, a projected effective Hamiltonian $\tilde H_\text{eff}$ can be constructed allowing for tractable dynamics within the projective subspace (bottom).}\label{fig:basisrotation}
	\end{figure}
	
	Finally, within the context of quantum dynamics, given some initial wave function $|\psi(0)\rangle$, we wish to find time evolution with respect to $H$ using this reduced Hilbert space. This is equivalent to
	\begin{align}
	|\psi(t)\rangle &= \exp\big(-itH\big)|\psi(0)\rangle \nonumber\\
	&=U\exp\big(-it\tilde{H}\big)|\tilde\psi(0)\rangle \nonumber \\
	&\approx U\exp\big(-it\tilde H_\text{eff}\big)|\tilde \psi(0)\rangle,
	\end{align}
	with $|\tilde \psi(0)\rangle = U^\dagger|\psi(0)\rangle$, and where the last expression is exact provided the initial wave function lies within the low-energy sector $\mathcal P|\tilde \psi(0)\rangle = |\tilde \psi(0)\rangle$. Expectation values of observables can be obtained in the standard way as

	\begin{align}
	\langle \psi(t)|\mathcal O|\psi(t)\rangle &= \langle \tilde \psi(0)|e^{it\tilde H}U^{\dagger}\mathcal O Ue^{-it\tilde H}|\tilde \psi(0)\rangle\nonumber\\
	&\approx \langle \tilde \psi(0)|e^{it\tilde H_{\text{eff}}}U^{\dagger}\mathcal O Ue^{-it\tilde H_{\text{eff}}}|\tilde \psi(0)\rangle. \label{eq:SW_tilde_dynamics}
	\end{align}

Thus, dynamics of the system can be obtained in two ways, as represented in Fig.~\ref{fig:basisrotation}. One can calculate dynamics with respect to $H$, which may be difficult due to an excessively large Hilbert space size. Alternatively, one may rotate and project all operators and observables to the transformed (``$\sim$") basis and calculate projected dynamics (as in Eq.~\eqref{eq:SW_tilde_dynamics}).

The main difficulty in the described above way of implementing the SW transformation is finding the rotation $U$, especially if the coupling $\lambda$ is not too small. Various perturbative expansions exist \cite{Bravyi2011a,Michailidis2018a}, but they mainly rely on the massive degeneracy of the ground state and large energy gaps in $H_0$ in order to be practical. There is an alternative approach based on first mapping the static Hamiltonian to the Floquet one, and then using the high-frequency expansions~\cite{Bukov2016a}, with similar complications of being generally asymptotic and uncontrolled unless taking the limit $\lambda\to 0$. A central result of this paper is developing a controllable and convergent, at least in principle, \textit{approximation} to the unitary $U$, which can be practically used even at intermediate values of the coupling $\lambda$.

	\subsection*{Generating the rotation}
	
	Rather than immediately calculating the unitary $U$, we will first compute its generator. Consider a family of unitary transformations $U(\mu)$ with $U(0)=\mathbbm{1}$ defined with respect to the running parameter $\mu\in 0\dots\lambda$ and their infinitesimal generators $A(\mu)=i  [\partial_\mu U(\mu)]U^\dagger(\mu)$ such that 
	
	\begin{equation}\label{eq:unitary_definition}
	    U^\dagger(\mu) = \mathcal{T}\exp\bigg(i\int^\mu_0 A(\mu')d\mu'\bigg),
	\end{equation}
	
where $\mathcal T$ stands for the path ordering symbol with respect to $\mu'$. At these intermediate points $\mu\in 0\dots\lambda$, one may define a parameterized Hamiltonian $H_0+\mu V$ which is rotated by the unitary $U(\mu)$ into the ``$\sim$" frame

	\begin{align}
	\tilde{H}(\mu) &= U^{\dagger}(\mu)\left(H_0+\mu V\right)U(\mu).
	\label{eq:rotatedHam}
	\end{align}
	The generator $A(\mu)$ is chosen such that at all points $\mu\in 0\dots \lambda$, the rotated Hamiltonian $\tilde H(\mu)$ is block diagonal in $P$ and $Q$,
	\begin{equation}\label{eq:intermediate_mu_blockdiag_condition}
	    \mathcal P \tilde H(\mu) \mathcal Q = \mathcal Q \tilde H(\mu) \mathcal P = 0.
	\end{equation}
	The unitary $U(\lambda)$ found in this way then generates the desired Schrieffer-Wolff rotation. To compute the form of $A(\mu)$ one may differentiate Eq. \eqref{eq:intermediate_mu_blockdiag_condition} with respect to $\mu$,
	\begin{align}
	&\partial_{\mu}\big[\mathcal P \tilde{H}(\mu)\mathcal Q\big]=0,\nonumber\\
	&\Rightarrow \mathcal P U^{\dagger}(\mu)\Big(V + i [A(\mu),H(\mu)]\Big) U(\mu)\mathcal Q=0.\label{eq:tilde_H_specific_generator}
	\end{align}
	 The solution to this equation is obviously not unique, since one can perform arbitrary unitary rotations within the sub-blocks $P$ and $Q$ as well as add any operator to $A(\mu)$ which commutes with $H(\mu)$.

\begin{figure*}
	    \centering
	    \includegraphics[width=0.9\linewidth]{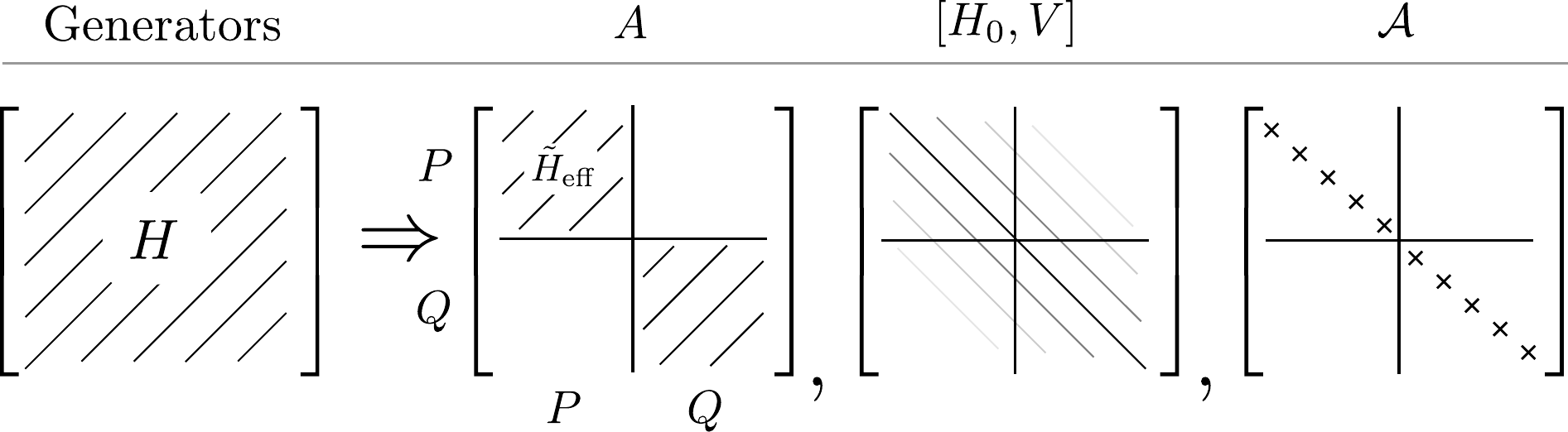}
	    \caption{Graphical representation of various transformations depending on the generator. A Hamiltonian $H$ is written in the eigenbasis of $H_0$ (which can be separated in subspaces $P+Q$), and generically has off-diagonal elements. The Schrieffer-Wolff generator $A$ block-diagonalizes this Hamiltonian, the Wegner-flow generated by the first-order approximation to the gauge potential $[H_0,V]$ band-diagonalizes this Hamiltonian by suppressing off-diagonal elements between states with large energy differences, and the exact adiabatic gauge potential $\mathcal A$ goes a step further by exactly diagonalizing $H$ in the eigenbasis of $H_0$.}
	    \label{fig:block_diagonal_diagram}
	\end{figure*}

    A particular solution to this equation is the adiabatic gauge potential (AGP) $\mathcal{A}(\mu)$ \cite{Kolodrubetz2017}, which is defined as the generator of evolution of instantaneous Hamiltonian eigenfunctions in parameter space: for Hamiltonian $H(\mu)$ and eigenstates $|n(\mu)\rangle$ the AGP $\mathcal{A}(\mu)$ satisfies
    \begin{equation}\label{eq:Adef}
        \mathcal{A}(\mu)|n(\mu)\rangle\equiv i\partial_\mu|n(\mu)\rangle.
    \end{equation}
    If $A(\mu)=\mathcal{A}(\mu)$, the rotated Hamiltonian $\tilde H$ becomes diagonal in the eigenbasis of $H_0$,
    \begin{align}
        \tilde H(\mu)=\sum_n&|n(0)\rangle E_n(\mu)\langle n(0)|.
    \end{align}
with the correct eigenvalues of $H_0+\mu V$. Thus, the AGP satisfies a stronger requirement than Eq.~\eqref{eq:intermediate_mu_blockdiag_condition} imposes: $\tilde H(\mu)$ has no off-diagonal matrix elements, not just those for states belonging to different subspaces. In other words, it is a good SW generator for \textit{any} choice of energy subspace $P$, $Q$. This fact is illustrated in Fig. \ref{fig:block_diagonal_diagram}. Recasting Eq. \eqref{eq:Adef} for all states \cite{Sels2016b} yields an operator equation similar to Eq. \eqref{eq:tilde_H_specific_generator} but without any projectors:
	\begin{equation}\label{eq:Adef2}
	\big[V + i[\mathcal{A}(\mu),H(\mu)],H(\mu)\big]=0.
	\end{equation}
This equation is even more difficult to solve than Eq.~\eqref{eq:tilde_H_specific_generator} because solving it amounts to fully diagonalizing the Hamiltonian. Moreover, the exact AGP is generally an exponentially divergent and highly nonlocal operator in the thermodynamic limit~\cite{Kolodrubetz2017}. However, there has been recent work on finding various \textit{local approximations} to the exact gauge potential \cite{Sels2016b, Claeys2019a}. Such approximations were shown to be able to efficiently reproduce the action of the gauge potential between states that can be distinguished by local operators. These can be states corresponding to either different energy sectors or symmetry sectors of $H_0$. In particular, local approximations of the AGP were shown to efficiently suppress matrix elements of the rotated Hamiltonian between states separated by large energies while failing to diagonalize it with states close in energy \cite{Claeys2019a}. As we will show below, the local AGP is also efficient at suppressing matrix elements between states with close energies as long as they belong to different blocks of the unperturbed Hamiltonian. Therefore identifying the generator of the SW transformation $A(\mu)$ with local approximations of the AGP leads to an accurate approximation to the unitary $U(\lambda)$.

	Another advantage of using the approximate AGP is that it can be found non-perturbatively, e.g. using a variational approach. This means the accuracy of the approximation is determined not by smallness of the perturbation of $\lambda$ but by the locality of rotations needed to block-diagonalize the Hamiltonian. It is also controlled by the size of the variational manifold used to find the approximate gauge potential: as the number of variational parameters increases the variational AGP approaches the exact one. As we will show using two particular examples, one can get a very good convergence even well beyond the regime of applicability of conventional perturbative approaches. More specifically, Eq.~\eqref{eq:Adef2} can be recast as the minimization of an action. Following \cite{Sels2016b}, suppose some ansatz for an approximate gauge potential over some subset of local operators $\{B_i\}$ (for example, all operators with a fixed finite spatial support)
	
	\begin{equation}
	    A(\mu,\{\alpha\}) = \sum_i \alpha_i B_i.
	\end{equation}
	
	One can compute the best variational solution to coefficients $\{\alpha_i\}$ by computing the minimum Hilbert-Schmidt tracenorm of Eq. \eqref{eq:Adef2}
	
	\begin{equation}
\label{eq:min_norm}
	    \text{MIN:}\quad\bigg|\bigg|\big[H(\mu),V + i[A(\mu,\{\alpha\}),H(\mu)]\big]\bigg|\bigg|.
	\end{equation}

	Finding this minimum amounts to quadratic minimization of coefficients $\alpha_i$ and the absolute minimum of this norm is achieved precisely when $A(\mu,\{\alpha\})=\mathcal A(\mu)$. Remarkably, since the resulting tracenorm can generally be calculated without constructing the operator in the full Hilbert space, the semi-analytic nature of such variational approaches allows for calculation of $A(\mu)$ in the thermodynamic limit (see Appendix \ref{appendix:finite_rotations}).

The last step consists of projecting the rotated Hamiltonian to the subspace. Given a projector on the subspace as $\mathcal P = \sum_{p\in\mathcal P}\ket{p}\bra{p}$, computing $H_{\rm eff}=\mathcal P \tilde H \mathcal P$ amounts to computing the matrix elements of the Hamiltonian $\tilde H$ between the states within the subspace $P$ since $\braket{p|\tilde H_{\rm eff}(\lambda)|p'}=\braket{p|\tilde H(\lambda)|p'}$, $p,p'\in P$. Similarly one can compute the matrix elements of all other observables. This procedure is feasible if the basis elements are well defined: for example, all elements in some symmetry sector, or low-energy eigenstates of an integrable model.

On passing we note that while we only analyze quenches and ground state response functions for time independent Hamiltonians, our approach can easily be extended to arbitrary protocols where the parameter $\lambda$ is time dependent, provided that this time dependence does not lead to strong coupling between blocks, e. g. via resonant transitions. Then one has to perform a time-dependent unitary rotation with $U(\lambda(t))$, which can be readily done as we are effectively computing the whole family of the unitaries $U(\mu)$ in some interval of $\mu$. Likewise we have full access to the term $-\dot \lambda A_\lambda$, which has to be added to the Hamiltonian transformed by a time-dependent unitary~\cite{Kolodrubetz2017}.

	\subsection*{Method implementation}
	Before illustrating the power of our approach we will summarize its implementation step by step.
	
	\begin{enumerate}
	    \item Find approximate generators along the range of $\mu\in [0,\lambda]$ by minimizing the Hilbert-Schmidt tracenorm of Eq. \eqref{eq:Adef2} for $A(\mu)$ constructed within a given (local) operator basis (see Refs.~\cite{Sels2016b, Claeys2019a} and discussion below for examples of possible basis choices and Appendix \ref{appendix:variational_minimization} for numerical implementation). 
	    
	    \item Compute rotated operators $\tilde{\mathcal O}\equiv \tilde{\mathcal O}(\lambda)=U^\dagger(\lambda)\mathcal OU(\lambda)$ including the Hamiltonian $H=H_0+\lambda V$ and observables. This may be done efficiently by computing Heisenberg evolution of some intermediary operator $Q(\mu)=U(\mu)U^\dagger(\lambda)\mathcal{O} U(\lambda)U(\mu)^\dagger$, which satisfies the following equation of motion:

        \begin{equation}\label{eq:Q_evolution}
	        \partial_{\mu}Q(\mu)=i[Q(\mu), A(\mu)].
	    \end{equation}
	    
	     By construction, $Q(\lambda)=\mathcal O$ as $UU^\dagger = \mathbbm{1}$, and $Q(0)=\tilde O$ as $U(0)=\mathbbm{1}$. Thus, this evolution should be propagated starting at $\mu=\lambda$ and evolving from $\mathcal Q(\lambda)=\mathcal O$ to $Q(0)=\tilde{\mathcal O}$. Critically, because the generator is local by construction, this evolution can be performed efficiently by taking advantage of locality as detailed in Appendix \ref{appendix:finite_rotations}.

	    \item Find the effective Hamiltonian $H_{\rm eff}$ by evaluating the matrix elements of $\tilde H$ in the original (unrotated) basis of $P$:
		\begin{equation*}	    
	    \braket{p|\tilde{H}_{\text{eff}}|p'} \equiv \braket{ p|\tilde{H}|p'},
	    \end{equation*}
	     for all states $p,p' \in P$. Diagonalizing $\tilde H_{\rm eff}$ will automatically generate the spectrum of the Hamiltonian within the dressed subspace $\tilde{P}$. If the latter stands for the low-energy sector, then this diagonalization yields approximate (dressed) eigenstates and eigenenergies of the ground state and low-energy excitations of $H$.

	    \item For describing quenches, i.e. evolution with a time-independent Hamiltonian starting from the initial state $|\psi\rangle$ belonging to the dressed subspace $\tilde{P}$, the rotated initial state is obtained by solving the Schr\"odinger equation again backwards in $\mu$: $i\partial_\mu \ket{\Psi(\mu)} = A(\mu)\ket{\Psi(\mu)}$ from $\mu= \lambda$ to $\mu=0$, with  $\ket{\Psi(\lambda)} = \ket{\psi}$ and $\ket{\Psi(0)} = \ket{\tilde{\psi}}$. This evolution can be calculated using e.g. Krylov methods or matrix product states \cite{Haegemann2011}, as it requires evolution in the full Hilbert space \textit{before} projecting to the subspace. 
	    
	    \item Time evolution of the observable $\mathcal O$ can now be calculated within the subspace $\tilde P$ by first evolving the wave function $\ket{\tilde{\psi}(0)}=\ket{\tilde{\psi}}$ with the Hamiltonian $\tilde{H}_{\text{eff}}$, leading to the time dependent state $\ket{\tilde{\psi}(t)}$ as
\be
\ket{\tilde{\psi}(t)}=\exp[- i \tilde H_{\rm eff} t] \ket{\tilde{\psi}(0)},
\ee
and second computing the expectation values of rotated observables such that
\be
\langle \psi(t)|\mathcal O |\psi(t)\rangle\approx
\braket{\tilde{\psi}(t)|\tilde{\mathcal{O}}|\tilde{\psi}(t)}.	
\ee     

    If the initial state $\ket{\psi}$ \textit{does not} belong to a single subspace $\tilde P$ then one has to first project this state into different subspaces $\tilde P_1, \tilde P_2,\dots$ and then evolve it separately in each subspace together with each subspace's effective Hamiltonian and observables. Because the Hamiltonian $\tilde H$ is approximately block diagonal, the wave function in each subspace evolves in time independently.
	\end{enumerate}

\subsection*{Connection with Wegner flow and perturbative SW transformations}
	As mentioned previously, it was recently argued by some of us that an accurate approximation to $A(\mu)$ can be found through a commutator expansion \cite{Claeys2019a}:
	\begin{equation}\label{eq:commexp}
    A(\mu,\{a\}) = i \sum_{k=1}^{\ell} a_k \underbrace{[H(\mu),[H(\mu),\dots [H(\mu)}_{2k-1}, V]]],
    \end{equation}
    where $\{a\}=\{a_1, a_2, \dots, a_{\ell}\}$ follows from the variational minimization.  Truncating this expansion to a single commutator level $(\ell=1)$, we get
    \begin{equation}
    \partial_{\mu}H(\mu) =  i a_1 [[H,V],H].
    \end{equation}
	This equation is highly reminiscent of the Wegner flow (also known as the similarity renormalization group \cite{Szpigel2000a}), where a flow equation is constructed for the Hamiltonian as $\partial_s H(s) = [\eta(s),H(s)]$, with the goal of obtaining a diagonal matrix for $s\to \infty$. A commonly-used generator is given by $[H(s),V(s)]$, where $V$ is the off-diagonal part of $H(s)$. This flow systematically suppresses off-diagonal elements of $H(s)$ in the same vein as the Schrieffer-Wolff generator, and it can be seen that a similar equation can be obtained by rescaling $\mu$ by $a_1$, with the crucial difference that the flow equation for the SW transformation only ranges in the interval $\mu\in[0,\lambda]$, whereas the Wegner flow necessitates the limit $s \to \infty$. This observation then also suggests that convergence of the Wegner flow may be improved by adding higher-order variationally-optimized commutators to the flow generator.

	Finally, we point out that the standard perturbative SW transformation is obtained when we approximate $U\approx \exp[-i \lambda A(0) ]$, where $A(0)$ is the solution of Eq.~\eqref{eq:Adef2} at $\lambda=0$, which simplifies to
	\begin{equation}
	[V+i[A(0),H_0],H_0] = 0.
	\end{equation}
This equation is exactly the one to be solved for defining the leading order in standard Schrieffer-Wolff transformation~\cite{Bravyi2011a}. Note that in some cases it is also possible to go to higher-orders of the perturbation in SW theory, which is equivalent to the van Vleck high-frequency expansion of the rotating frame Floquet Hamiltonian~\cite{Bukov2016a}. Then it is possible to systematically develop perturbation theory for the generator of the SW transformation. Usually, such perturbation expansions are more computationally demanding than the variational method.

	\subsection*{Example}
	For an explicit example where the first-order commutator expansion is exact, let us choose a Hamiltonian as in Fig.~\ref{fig:SW_schematic},
    \begin{equation}
        H = H_0 + \lambda (V_+ + V_-),\quad V_+^\dagger=V_-,
\label{eq:first_example}
    \end{equation}
    where $H_0$ consists of degenerate levels separated by $\Omega$, and $V_\pm$ acting on an eigenstate of $H_0$ can only change the energy by $\pm\Omega$. This requirement leads to commutation relations $[H_0,V_\pm]=\pm\Omega V_\pm$. While such a model might seem somewhat artificial, such Hamiltonians are commonly encountered in Floquet systems~\cite{Bukov2019} and standard SW transformations (see e.g. the Fermi-Hubbard model below).
    
    Considering the expansion from Eq.~\eqref{eq:commexp} and keeping only the first-order term leads to 
    \begin{equation}
        A(\mu) = ia_1(\mu)[H,V] =ia_1(\mu)\Omega\big(V_+-V_-\big).
    \end{equation}
    Plugging this equation in the variational minimization for $\mu=0$ results in finding the minimum of
    \begin{align}
    &\tr{(V_++V_--a_1(0) \Omega [V_+-V_-,H_0])^2} \nonumber\\
    &\qquad = (1+a_1(0) \Omega^2)^2\tr{(V_+ + V_-)^2}.
    \end{align}
    This expression is exactly zero when $a_1(0)=-1/\Omega^2$, leading to $A(0)=-i(V_+-V_-)/\Omega$. For small $\lambda/\Omega$ the rotation can be expanded up to $\mathcal{O}(\lambda^2)$ to return
    \begin{eqnarray}
    U^\dagger H U&\approx&H + [A,H] + \frac{1}{2}\big[A,[A,H]\big]\\
    &=&H_0 - \frac{\lambda^2}{\Omega}\big[V_+,V_-\big].
    \end{eqnarray}
	The first commutator exactly cancels the perturbative term, yielding a new Hamiltonian which is block-diagonal in the eigenbasis of $H_0$ up to $\mathcal{O}(\lambda^2)$, returning the standard SW results. In order to go beyond these results, it is possible to retain the higher-order terms in the commutator expansion and perform the rotation in a more involved way, which is precisely explained below for specific models.
 
    In general, the critical parts of the variational commutator expansion are two-fold. First, the magnitude of the rotation is controlled by the size of gaps $\Omega$ in the system, but does not necessarily require any additional structure in the original Hamiltonian: for example, it could have a varying Hilbert subspace size, not have exact gap differences, or not be degenerate within each subspace. These are significant relaxations on traditional Schrieffer-Wolff transformations, which normally require exact degeneracies in order to be feasible.
    
	The second advantage of these expansions is that it is by nature local; the $n$-th order term of the expansion will have operator support on the order of $n$ sites. This means, for order-1 parameter $\lambda/\Omega$, the rotated wave function $|\tilde \psi\rangle$ is only entangled within some finite support and the Hamiltonian $\tilde H$ is similarly quasi-local, due to locality of $A(\mu)$ and bounded evolution ``time'' in the $\lambda$-space. With this observation in hand, existing methods which take advantage of locality may be readily applied to extract the basis-rotated objects, as also detailed in Appendix~\ref{appendix:finite_rotations}.

	\section{Fermi-Hubbard Model}
	
	A classic example to apply the proposed method to is the attractive Fermi-Hubbard model, where the perturbative Schrieffer-Wolff transformation returns the well-known Heisenberg model \cite{Hubbard1963,Shastry1986b,Shastry1986c}. Here, this model will be used to illustrate the non-perturbative nature of the variationally-obtained rotations, which allows for effective dynamics at values of $\lambda/\Omega$ where the perturbative SW transformation is no longer expected to return accurate results. The disordered Fermi-Hubbard Hamiltonian is given by
	\begin{align}
	H =  &- \Omega\sum_i \left(n_{i,\uparrow} -{1\over 2}\right)\left(n_{i,\downarrow} -{1\over 2}\right)	+\sum_i\delta_{i\sigma}n_{i,\sigma}\nonumber\\
	&+\lambda \sum_{i,\sigma} \left(c^\dagger_{i,\sigma}c_{i+1,\sigma} -h.c.\right),
\label{eq:FHM}
	\end{align}
	in which $n_{i,\sigma} = c^{\dagger}_{i,\sigma}c_{i,\sigma}$ and $\delta_i$ are independent random numbers, which are drawn from a normal distribution with variance $\lambda^2 \Delta/2\Omega$. The factor $\lambda^2/2\Omega$ is such that the crossover from weak to strong disorder regimes in the effective Heisenberg model happens at $\Delta~\sim 1$.
	For $N$ sites, there are of the order of $4^N$ degrees of freedom, with 4 possible fermion states for each site $\{\ket{0},\ket{\uparrow},\ket{\downarrow},\ket{\uparrow \downarrow}\equiv\ket{2}\}$.

	Given the expansion from the $\lambda=0$ point we choose $H_0$ to be the Hubbard interaction and disorder term and $V$ to be the nearest-neighbor hopping terms. Notice that the disorder breaks the otherwise massive degeneracy of the unperturbed Hamiltonian $H_0$. The hopping can change energy of $H_0$ by either $\pm\Omega$ if it corresponds to changing the number of singly occupied sites (spinons),  or by 0 if it conserves this number. For this reason the Fermi-Hubbard model is a more complicated version of the Hamiltonian~\eqref{eq:first_example}, though it shares many of its features. We choose the commutator expansion~\eqref{eq:commexp} to get the first two basis operators of the variational gauge potential
	\begin{multline}
	A(\mu) = i a_0(\mu)[H(\mu),V] \\
+i a_1(\mu)\big[H(\mu),\big[H(\mu),[H(\mu),V]\big]\big].
	\end{multline}
Due to the local structure of the Hamiltonian, this approximate generator is also local, constrained to operators with span of 4 sites or less. Specifically, the first commutator has 12 terms per site and appears as
	\begin{align}
	[H(\mu),V]=&\sum_{i,\sigma}\bigg(\frac{\delta_i}{\Omega} +n_{i,\overline \sigma} - n_{i+1,\overline \sigma}\bigg)\nonumber\\
	&\qquad \times \bigg(c_{i, \sigma}^\dagger c_{i+1,\sigma} + c_{i,\sigma}c^\dagger_{i+1,\sigma}\bigg).
	\end{align}
	The second commutator has 144 terms per site, which is intractable to write down and solve by hand; instead we turn the computer to calculate the commutators. 

Next, we minimize the norm in Eq.~\eqref{eq:min_norm} within the operator space spanned by these two commutators, which is equivalent to inversion of a $2\times 2$ matrix to get a variational approximation of $A(\mu)$. We choose to compute the AGP in the disorder-free case for computational simplicity, and apply it to the disordered Hamiltonian. While this may generate some error at large disorder, it is suppressed by powers of at least $(\lambda/\Omega)^3$: two orders from the base disorder strength of $\Delta\lambda^2/\Omega$, an additional power stemming from the fact that the unrotated operator is still block-diagonal. We checked numerically that in the analyzed regimes the variational gauge potential does not change significantly when computed with vs. without disorder and this difference only slightly affects the results presented below~\footnote{\label{footnote1} For $\Omega/\lambda=7.5$, 6 sites, and $\Delta=10$ the operator norm of the difference between the AGP computed with and without disorder is $||dA||\approx 0.05||A||$. The wave-function fidelity of a N\'eel state is changed by less then 0.5\%.}.

	Let us start in the perturbative regime $\lambda/\Omega\ll 1$. Here, one may compute the AGP in powers of $\mu$ as $A(\mu) = A_0 + \mu A_1$. Then we can reduce the generator of the SW rotation to a ``time"-independent operator via a Magnus expansion~\cite{Blanes2009}
    \begin{equation}
\mathcal T\exp\left(i{\int_0^\lambda A(\mu) d\mu}\right)\approx \exp\left(i\lambda A_0 + i\frac{\lambda^2}{2}A_1+\mathcal O(\lambda^3)\right).
    \end{equation}
The rotated Hamiltonian can then be computed via a second-order BCH expansion \cite{Griffiths2005}, keeping terms up to order $\lambda^2$
	\begin{align*}
	\tilde H=-& \Omega\sum_i \left(n_{i,\uparrow} -{1\over 2}\right)\left(n_{i,\downarrow} -{1\over 2}\right)+\delta_i(n_{i,\uparrow} + n_{i,\downarrow})\\
	+&\lambda\sum_{i,\sigma}\big(1-(n_{i+1,\overline\sigma} - n_{i,\overline\sigma})^2\big)\big(c^\dagger_{i,\sigma}c_{i+1,\sigma} - c_{i,\sigma}c^\dagger_{i+1,\sigma}\big) \nonumber\\
    -&\frac{\lambda^2}{\Omega}\sum_{i,\sigma}(c_{i,\sigma}c_{i,\overline\sigma})(c^\dagger_{i+1,\sigma}c^\dagger_{i+1,\overline\sigma}) + \text{c.c.}\\
    +&\frac{\lambda^2}{\Omega}\sum_{i,\sigma}(c_{i,\sigma}c^\dagger_{i,\overline\sigma})(c^\dagger_{i+1,\sigma}c_{i+1,\overline\sigma}) + \text{c.c.}\\
    +&\frac{\lambda^2}{\Omega}\sum_{i\sigma}(1-2n_{i,\overline\sigma})(1-2n_{i+1,\sigma}).
	\end{align*}
Here, the blocks diagonalizing the leading order-$\Omega$ term define the subspaces, while the order-$\lambda$ term describes hopping of spinons and doublons: the $1-(n_{i+1,\overline\sigma} - n_{i,\overline\sigma})^2$ term suppresses hopping between states belonging to different subspaces. The order-$\lambda^2/\Omega$ terms describe either the double hopping between adjacent sites (the third term) or spin exchange between adjacent spinons (fourth term).
	
	This Hamiltonian still acts on the full Hilbert space but it is block diagonal up to $\lambda^2/\Omega$. For this reason the projective subspace $P$ can be chosen to be any sector of fixed number of singly-occupied sites (spinons). One such choice is that of no singly occupied sites, which corresponds to the lowest-energy subspace for $\Omega>0$. The Hamiltonian confined to this subspace can be in turn mapped to a spin Hamiltonian by identifying $\ket{0} \to \ket{\Downarrow}$ and $\ket{2} \to \ket{\Uparrow}$, as well as operators $(c_{i,\sigma}c_{i,\overline\sigma})\to \sigma_i^-$ and so forth. The effective Hamiltonian in this subspace is then given by the Heisenberg model.
	
	In general, the subspace $P$ does not necessarily need to be the lowest-energy subspace, just one of the energy blocks of $H_0$. In this way, one may compute the effective Hamiltonian with any number of spinons as well. Note that in those sectors the hopping term will be non-zero and thus the time evolution there will be dominated by a timescale $\lambda$ corresponding to spinon hopping. Because the rotated Hamiltonian is block-diagonal, each subsector can be time-evolved independently.

	The above derivation was perturbative, in that the rotation was computed using a perturbative BCH expansion, which requires careful power-counting. It comes as no surprise that it is asymptotically exact in the limit $\lambda/\Omega \to 0$: the hopping terms raise or lower the energy by $\pm\Omega$, and so this model is a more explicit version of the example~\eqref{eq:first_example}. In principle, one could go to higher orders of perturbation theory or SW to compute corrections, but in practice it becomes unwieldy. Instead, this same process can be applied for finite $\lambda$ by computing the generator $A(\mu)$ at each step $\mu$ variationally, then computing the rotated operators numerically (see Appendix). These numerical computations take advantage of the fact that the generator is local. In this way, the only error comes from the variational approximation for the gauge potential, whose validity can be found by increasing the variational ansatz size.

	\begin{figure}
		\includegraphics[width=\linewidth]{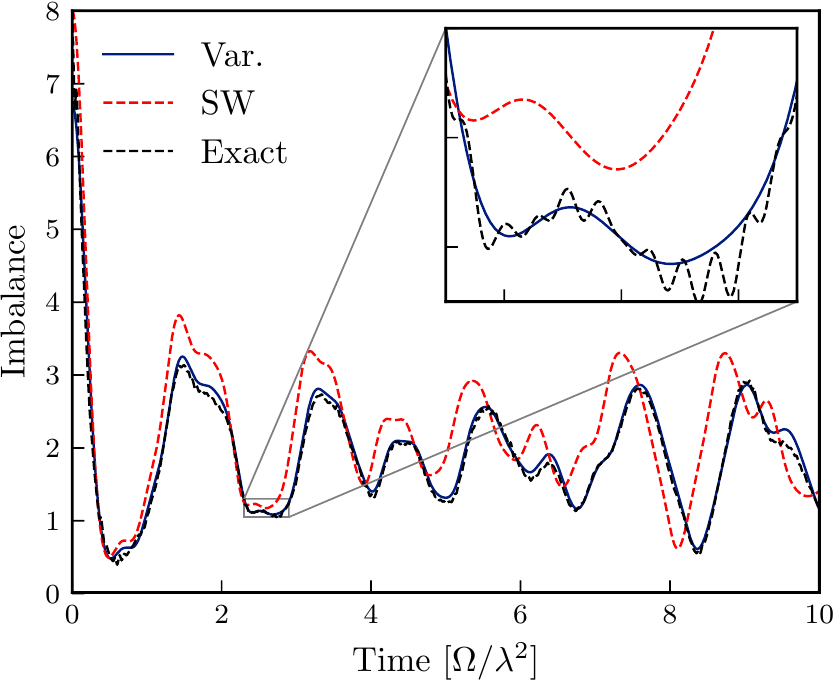}
		\caption{Comparison exact (black dashed line), projected variational (blue solid line), and perturbative SW (red dashed line) dynamics of the Fermi-Hubbard model for a small system of 8 fermionic sites, $\Omega/\lambda=5.0$ and a single disorder realization of strength $\Delta=2.5$. The initial condition is a N\'eel state of alternating doubly-occupied and non-occupied sites, with periodic boundary conditions. Inset details how the projected dynamics miss high-frequency oscillations.}\label{fig:fermi-hubbard-compare}
	\end{figure}

	\begin{figure}
		\includegraphics[width=\linewidth]{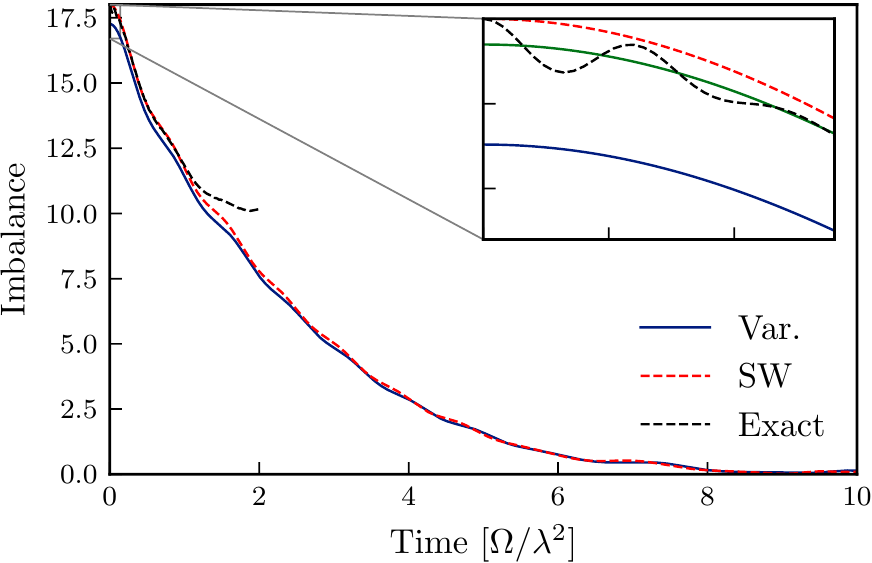}
		\caption{Comparison of exact and projected dynamics of the Fermi-Hubbard model for a large system of 18 fermionic sites, beyond the reach of exact dynamics, with $\Omega/\lambda=5$ and a boundary-wall initial condition. The black dashed line represents exact results for a small system of 8 sites, where finite-size effects occur after a time of order $2$. Error at time $t=0$ is due to missing overlap w.r.t. higher-energy sectors. This can be seen from the green line in the inset, representing the variational results renormalized by the fidelity of the projected initial state as $\langle \tilde O(t)\rangle / \langle \tilde{\mathcal{P}} \rangle$. }\label{fig:fermi-hubbard-boundary-wall}
	\end{figure}

    \subsection*{Quench Dynamics of the Fermi Hubbard Model}

    We will now analyze quench dynamics of the Fermi-Hubbard model starting from N\'eel and boundary-wall initial states. We choose the subspace $P$ as the space containing no singly-occupied sites, and thus would recover the Heisenberg model for $\lambda\to 0$.
    
    In Fig.~\ref{fig:fermi-hubbard-compare} we show a comparison of the fermion imbalance computed by a variationally-projected model, the perturbative Schrieffer-Wolff model, i.e. the Heisenberg model with disorder, and the original Hubbard model, for a small system of $8$ sites. The initial condition is chosen to be a N\'eel state of doubly-occupied sites $\ket{\psi(0)} = \ket{20202020}$. The Hamiltonian is chosen to have the ratio of the Hubbard interaction and the hopping $\Omega/\lambda=5$ and disorder strength $\Delta=2.5$. We calculate the density imbalance as the expectation value of $\mathcal I = \sum_{i,\sigma} (-1)^i n_{i,\sigma}$, which is extremal at $t=0$ and is expected to vanish if the system thermalizes.
    
    To demonstrate applicability of the method to go well beyond system sizes amenable to exact diagonalization,  in Fig.~\ref{fig:fermi-hubbard-boundary-wall} we show the imbalance for a quench from a domain wall initial condition of 18 sites for $\Omega/\lambda=5$ and zero disorder, where
     \begin{align}
	\ket{\psi(0)} =&\ket{222222222000000000 }.
	\end{align}
    Computing the exact time evolution requires access to approximately $4^{18}$ degrees of freedom (36 qubits), which is on the edge of computational feasibility, although variational methods such as DMRG may perform well. For this plot and initial state the relevant imbalance is defined as $\mathcal I = \sum_{i=1}^9 n_i - \sum_{i=10}^{18} n_i$, again maximal at the initial time $t=0$ and vanishing in time as the system thermalizes and the boundary wall dissolves.

The results illustrated in Figs.~\ref{fig:fermi-hubbard-compare} and~\ref{fig:fermi-hubbard-boundary-wall} highlight two important aspects of the method.  Fig.~\ref{fig:fermi-hubbard-compare} demonstrates that the method can be used to go beyond standard perturbative SW transformations and give a systematic and significant improvement to the perturbative results.  At the same time Fig.~\ref{fig:fermi-hubbard-boundary-wall} shows that the domain wall dynamics is very accurately described by the Heisenberg model, even in the regime where such an accuracy might not be anticipated.  As can be seen from the inset of Fig.~\ref{fig:fermi-hubbard-compare} the variational method accurately reproduces the low-frequency behavior, while failing to reproduce the high-frequency oscillations. These high-frequency oscillations originate from the fact that the dynamics follow a sudden quench, which excites the initial wave function beyond the lowest block. A more realistic and experimentally relevant situation is a gradual ramp of the coupling, which might be still fast with respect to the effective low-energy degrees of freedom, but slow with respect to the scale $\Omega$. These high-frequency oscillations are then expected to be strongly suppressed. In order to reproduce these fast oscillations within our scheme one needs to add evolution coming from other blocks, which can be done in parallel and hence does not significantly increase the complexity of the computation.

	\subsection{Wave function fidelity }
	
The quench effects also lead to a small mistake in the imbalance for an initial domain wall state even at the initial time, as illustrated in the inset of Fig.~\ref{fig:fermi-hubbard-boundary-wall}. As we explained above this happens because the wave function needs first be rotated, and then projected into the subspace; the rotation can result in a nonzero projection of the rotated wave function $|\tilde \psi\rangle$ to other subspaces even if the unrotated state $|\psi\rangle$ is fully contained in $P$. Mathematically this can be expressed in the partial loss of fidelity of the rotated wave function
	\begin{equation}
	    \langle \psi | U\mathcal{P}U^\dagger|\psi\rangle=\langle \tilde \psi|\mathcal P|\tilde\psi\rangle =\langle \psi|\tilde{\mathcal P}|\psi\rangle\leq 1.
	\end{equation}
	The subspace $P$ contains the lowest-energy eigenstates of the \textit{non-interacting} model, including the N\'eel and boundary wall states. However, the subspace $\tilde{P}$ can be seen as the lowest-energy eigenstates of the \textit{interacting} model, which may include mixed spin-charge degrees of freedom. Equivalently, a quench can be seen as injecting some finite energy density into the system, such that there must be some overlap with higher-energy sectors with some finite number of defects (spinons or doublons), which would be indicated here as a wavefunction fidelity less than one.
	
	This loss can be recovered by adding extra subspaces $P_i$ and evolving each independently, then resumming observables within each subspace:
\begin{equation}\label{eq:all_subspace_observable}
	    \langle O(t)\rangle \approx \langle \tilde\psi(t)|\tilde O|\tilde \psi(t)\rangle+\sum_n\langle \tilde\psi_n(t)|\tilde O_n|\tilde \psi_n(t)\rangle,
	\end{equation}
where the index $n$ denotes the wave function or operator within the higher subspaces $n$. For N\'eel/boundary wall quenches, these higher subspaces correspond to $2n$ singly-occupied sites.

	The fidelity in  the lowest energy sector can be well-described as the probability of having zero defects in the system. In the dilute limit these defects appear with independent probabilities $\rho(\lambda/\Omega)$ such that,
	\begin{equation}\label{eq:fidelity_scaling}
	    \big|\langle\tilde\psi|\mathcal P|\tilde \psi\rangle\big| \approx \bigg(1-\rho(\lambda/\Omega)\bigg)^n,
	\end{equation}
\ We numerically checked that for the initial N\'eel state $\rho(x)$ is well fitted by $\rho(x)\approx \frac{x^2}{2}-\frac{3x^4}{2}+\mathcal O(x^6)$ \footnote{This functional form is changed slightly by computing the AGP with disorder}. For $\Omega/\lambda=5$ this expression gives $\rho\approx 0.017$, i.e. approximately $1.7\%$ chance of exciting a spinon pair per site. The fidelity is thus exponentially suppressed in the system size, and thus a better thermodynamic description would generally correspond to a sector with a  small, finite density of spinons. Vanishing fidelity in the thermodynamic limit is of course related to the well-known orthogonality catastrophe \cite{Anderson1967}, which is often easy to forget about especially if the perturbative limit $\Omega/\lambda\to\infty$ is taken before the thermodynamic limit $N\to\infty$. In passing, we comment that this dressing may have similarities to polarons \cite{Grusdt2018,Koepsell2019}, which are particle excitations dressed by spin degrees of freedom through interactions. The rotation $U$ may serve the same purpose of dressing such purely particle excitations.
    
\begin{figure}[t]

\subfloat[System of 18 fermionic sites and varying disorder and $\Omega$. Inset is the imbalance rescaled by the fidelity, collapsing all lines to the $\Omega \to \infty$ result and indicating that the loss is due to overlap with thermalizing finite-spinon sectors.]{%
  \includegraphics[width=\columnwidth]{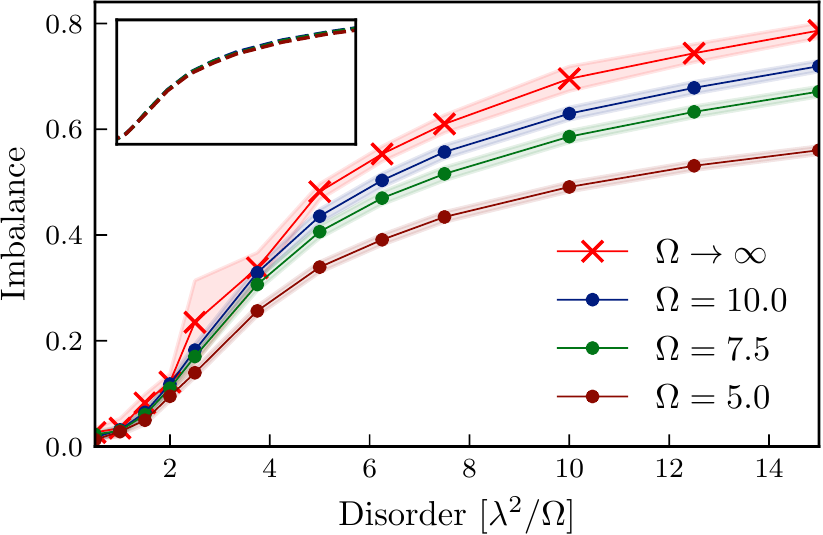}%
  \label{fig:fermi-hubbard-MBL}
}

\subfloat[Varying system size at fixed disorder $\Delta=10$ and $\lambda/\Omega=7.5$. Small system sizes are consistent with exact results (black).]{%
  \includegraphics[width=\columnwidth]{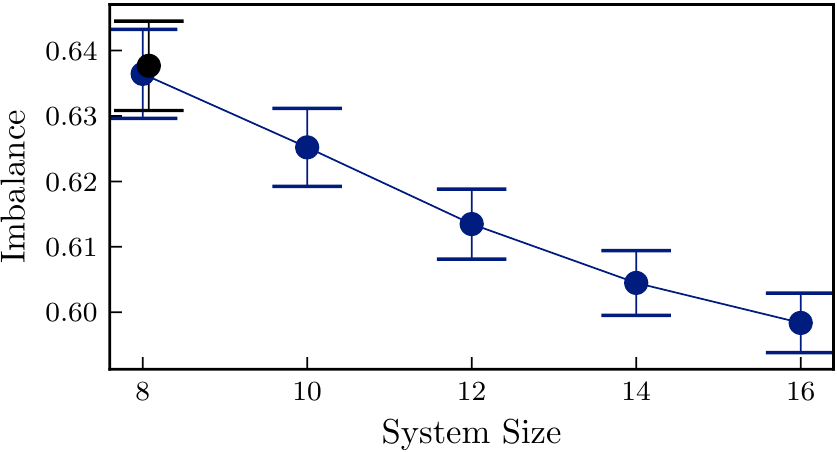}%
  \label{fig:fermi-hubbard-systemsizes}
}

\caption{Steady-state imbalance for a disordered Fermi-Hubbard model. The initial condition is a N\'eel state of alternating doubly-occupied and non-occupied sites with periodic boundary conditions, and is computed at time $t=25\Omega^2/\lambda$. The imbalance decreases with decreasing $\Omega$ and increasing system size, mainly due to fidelity loss.\label{fig:fermi-hubbard-MBL-full}}

\end{figure}

	\subsection{Many-body localization at finite $\Omega$}
	
We will now apply this method to analyze the effects of onsite disorder on the Fermi-Hubbard model. In the $\Omega\to\infty$ Heisenberg limit, it is widely believed that the model exhibits a many-body localization (MBL) transition indicated by long-time memory of initial conditions and lack of conductivity in equilibrium \cite{Gopalakrishnan2015,Agarwal2015}. The situation becomes much less clear when the ratio $\lambda/\Omega$ becomes finite and the mapping to the Heisenberg model starts to break down. In Fig.~\ref{fig:fermi-hubbard-MBL}, we present the results of simulations of the long-time  ($t=25\Omega^2/\lambda$) imbalance of an 18-site Hubbard chain with the N\'eel initial condition as a function of the disorder $\Delta$ and different ratios of $\Omega/\lambda$. Here a nonzero long-time value of the imbalance is an indicator of localization~\cite{Bloch2016,Abanin2019}. Note that the disorder strength in the Hubbard Hamiltonian is appropriately rescaled by a factor $ \lambda^2 /\Omega$ (c.f. Eq.~\eqref{eq:FHM}). As can be clearly seen, for smaller values of $\Omega/\lambda$ the late-time imbalance decreases, which can be heuristically explained by proliferation of spinons at smaller values $\Omega/\lambda$. The spinons hop at a much faster time scale than the spin exchange, allowing the otherwise MBL-frozen state to thermalize.
	
This qualitative reasoning can be quantified using the variational SW method developed here. We can check that the fidelity of the initial 18-site N\'eel state with the lowest energy subspace is $0.73$, $0.86$, and $0.92$ for $\Omega/\lambda=5$, $7.5$, and $10$, respectively, fully consistent with Eq.~\eqref{eq:fidelity_scaling}. Such relatively small numbers, especially for the lowest analyzed ratio $\Omega/\lambda=5$, imply that already for the system sizes studied there is a significant fraction of spinons present in the system. On top of that the effective Hamiltonian is also modified slightly: there are longer-range spin-spin interaction terms and emergent weak correlations in the disorder. However, we checked numerically (results are not shown here) that these effects are small and do not lead to thermalization within the subspace $\tilde P$, at least for these system sizes. This was done by computing ZZ time correlation functions within the $\tilde P$ sector. At the same time, the late-time imbalance shown in Fig. \ref{fig:fermi-hubbard-MBL} normalized by the fidelity as $\frac{\langle \tilde O(t)\rangle}{\langle \tilde{\mathcal{P}}\rangle}$ corresponds closely to the $\Omega\to\infty$ case, as shown in the inset of Fig. \ref{fig:fermi-hubbard-MBL}. Thus, the apparent decrease in imbalance is mostly due to projective loss of the wave function to other spinon sectors.

This loss can be compensated by analyzing the dynamics of these sectors and combining these results together. However, it turns out to be unnecessary as free spinons lead to a very rapid decay to zero of the imbalance because their hopping $\lambda$ is much larger than the disorder strength. Thus the sectors containing free spinons do not affect the long time imbalance as shown in Fig.~\ref{fig:fermi-hubbard-MBL}.  It can easily be shown that even a single spinon moving on top of a N\'eel state destroys the magnetic order~\cite{Shraiman1988}. For example, as illustrated below a spinon moving from right to left swaps the states $|0\rangle\leftrightarrow|2\rangle $ in the middle
	\begin{equation}\label{eq:normalized_observable}
	    \ket{2,0,2,\uparrow}\to\ket{2,0,\uparrow, 2}\to \ket{2,\uparrow, 0,2}\to \ket{\uparrow, 2,0,2}.
	\end{equation}
We confirmed these considerations by checking numerically that the higher-defect sectors always thermalize.
	
	The immediate implication of these observations is on the absence of MBL in the Fermi-Hubbard model. Indeed, for any finite $\lambda/\Omega$, there will be some density of defects, and the overlap with the zero-defect sector will be exponentially small in the system size, with exponential prefactor $\lambda^2/\Omega^2$. The zero-defect sector is believed to exhibit MBL behavior in the Heisenberg limit \cite{Nandkishore2015} (with some recent contestation \cite{Prosen2019}), which appears to be robust for finite $\lambda/\Omega$ (see above). However, due to inevitable spinon excitations, the long-time imbalance in the original Hubbard model always goes to zero in the thermodynamic limit. An example of this behavior is shown in Fig.~\ref{fig:fermi-hubbard-systemsizes}, where the imbalance is plotted as a function of system size for disorder $\Delta=10$ and $\Omega/\lambda=7.5$. The density of spinon excitations and hence the imbalance decay can be reduced by considering a smooth ramps of the hopping strength $\lambda$ instead of a quench or by going to larger ratios $\Omega/\lambda$. However, it is virtually impossible to eliminate spinons entirely, and thus care must be taken in finding the overlap of the rotated initial state $\ket{\tilde \psi}$ with the zero-spinon subspace.

	\section{The integrability-broken XY model}
	
	Another application of the Schrieffer-Wolff transformations we consider here is finding response functions of operators around the ground state $|\emptyset\rangle$ of some Hamiltonian $H$, such as
	\begin{equation}
	    C_{XY}(t)=\langle\emptyset |X(t)Y(0)|\emptyset\rangle.
	\end{equation}
	Response functions, and their Fourier-space counterparts structure functions, are fundamental objects describing low-energy excitations such as particles \cite{Sachdev2011a}. For local operators the energy of the wave function $Y|\emptyset\rangle$ is sub-extensive; a low-energy subspace of $H$ suffices to describe the dynamics of this response function as
	\begin{eqnarray}
	C_{XY}(t)&=&\langle \emptyset|X e^{-itH}Y|\emptyset\rangle \nonumber\\
	&\approx&\langle \emptyset|X\exp\bigg({-it\mathcal N H \mathcal N}\bigg)Y|\emptyset\rangle,
	\end{eqnarray}
	where $\mathcal{N}$ is the projection to $N$ low-lying eigenstates of the Hamiltonian $H$. However, computing these excited states for a generic interacting model is generally intractable, as one must deal with the exponential size of the basis set. As such, we turn to the Schrieffer-Wolff transformation to find some rotation $U$ to the ``$\sim$" basis, and then project the rotated Hamiltonian to a subspace with known analytic properties, avoiding the otherwise exponential complexity of the system. Inserting this rotation $UU^\dagger=\mathbbm{1}$ into the above leads to
	\begin{align}
	C_{XY}(t) \approx \langle \emptyset|X U\exp\bigg(-i t\tilde{\mathcal{N}} \tilde{H} \tilde{\mathcal{N}} \bigg)U^\dagger Y|\emptyset\rangle\nonumber.
	\end{align}
	The SW rotation can thus be recognized if a simple low-energy subspace $\mathcal P=\sum_{n\in N}|n(0)\rangle\langle n(0)|$ is associated with the interacting one as $U^\dagger\mathcal NU=\mathcal P=\tilde{\mathcal N}$. We further define $|\emptyset\rangle = U|\emptyset_0\rangle$, e.g. the interacting ground state rotated from the free basis. It is known that if there is no phase transition between Hamiltonians $H$ and $H_0$, there exists a quasi-local generator of rotations which maps between these ground states, exactly corresponding to the local gauge potential \cite{Bachmann2017a}. The response function can then be approximated as
	\begin{equation}
	C_{XY}(t)=\langle \emptyset_0|\tilde Xe^{-it\mathcal P \tilde H \mathcal P}\tilde Y|\emptyset_0\rangle,
	\end{equation}
	with the $``\sim"$ basis being the one generated by rotating using $U$. As seen previously, for a Hamiltonian $H = H_0+\lambda V$, the generator can be variationally obtained if the projective subspace are low-energy eigenstates of $H_0$, written as $\{|n_0\rangle\}$. Explicitly writing this projector as a sum, we find
	\begin{equation}
	    C_{XY}(t) = \sum_{nm}\langle \emptyset_0|\tilde X|n_0\rangle \langle m_0| \tilde Y|\emptyset_0\rangle \exp\big(-it\langle n_0|\tilde H|m_0\rangle\big).
	\end{equation}
	This is exact in the limit where the generator is the exact gauge potential, or when the subspace is the full space. Thus, the challenge of calculating response functions reduces to two tasks: (i) computing an appropriate generator of rotations, and (ii) computing matrix elements of operators. As argued before, task (i) can be performed variationally. Task (ii) is implementable, in principle, if the structure of eigenstates are known analytically. This is the case if the Hamiltonian $H_0$ is integrable and thus the eigenstates are written as particle excitations on top of some vacuum~\cite{Faddeev1996,Karbach1998}. The operators $\tilde X$ are quasi-local, as the generator of the rotations is local by the ansatz. This quasi-locality gives some hope of computing the matrix elements analytically, which would allow for the calculation of approximate dynamics even when the Hilbert space of the effective Hamiltonian becomes intractably large.

	Thus, this recipe will perform well for the following set of models. Given some integrable Hamiltonian $H_0$ and an integrability-breaking term $V$ with strength $\lambda$, one may compute low-energy excitations of the Hamiltonian $H=H_0+\lambda V$. One can anticipate that the integrability-breaking terms can even become non-perturbative, as long as the system is relatively far from any phase transition.	As an example, let us choose a relatively simple integrable system described by an XY-type model and an additional integrability-breaking term in the form of a longitudinal magnetic field,
	\begin{equation}
	H = \sum_i J_{xx}\sigma_x^i\sigma_x^{i+1} + J_{yy}\sigma_y^i\sigma_y^{i+1} + h\sigma_z^i + \lambda \sigma_x^i.
	\end{equation}\label{eq:U1_XYHamiltonian}
	For $\lambda=0$ this model maps to free fermions under a Jordan-Wigner transformation~\cite{Sachdev2011a}. For $J_{yy}=0$ this is the Transverse Field Ising Model, while for $J_{xx}=J_{yy}$ this is the XY model. For $\lambda=0$ the eigenstates can be written in terms of fermionic raising operators $\gamma_k^\dagger$ acting on some ground state, where each adds one particle of momentum $k$ to the system
	\begin{equation}
	|k,k', \dots ,k''\rangle = \gamma_k^\dagger \gamma_{k'}^\dagger\dots \gamma_{k''}^\dagger|\emptyset\rangle.
	\end{equation}
	The integrability-breaking term $\lambda \sigma_x^i$ in Jordan-Wigner notation can be seen as coherently adding and removing fermionic excitations from the system; it breaks the conservation laws in the system preserving the number of fermions. It also breaks the $\mathbb{Z}_2$ symmetry, as well as the $U(1)$ symmetry for the XY point. For the following example, let us choose periodic boundary conditions and parameters
	
	\begin{equation}
	    J_{xx}=J_{yy}=1,\ h=3,\ \lambda = 1.25.
	\end{equation}
	
	Here, the integrability breaking term is \textit{non-perturbative}, in the sense that it is of the same order as the other terms; there are no symmetries other than geometric ones such as translations, and the model is quantum chaotic (as shown in Appendix \ref{app:wignerdyson}). Of course, special eigenstates \cite{Bernien2017}
	such as those at the edges of the spectrum can preserve their integrable structure. Let us then proceed by calculating the generator $A(\mu)$ variationally. For this example, we will choose the variational manifold consisting of all operators with support up to three sites:
	\begin{eqnarray}
	A(\mu) &=& a_0^i\sigma_x^i + a_1^i\sigma_y^i+a_2^i\sigma_z+\nonumber\\
	&+&a_3^i\sigma_x^i\sigma_y^{i+1} + a_4^i \sigma_y^i\sigma_z^{i+1}+\dots\nonumber\\
	&+&a_5^i\sigma_x^i\sigma_y^{i+1}\sigma_z^{i+2} + a_6^i\sigma_z^i\sigma_y^{i+2},+\dots
	\end{eqnarray}
where all coefficients $a_j$ are $\mu$-dependent. This anatz gives a variational minimization procedure on $63N$ parameters, which have to be computed in the interval $\mu\in[0,\lambda]$. A further simplification comes from noting that for any real Hamiltonian the AGP is strictly imaginary so only the terms containing an odd number of $\sigma_y$ matrices are non zero~\cite{Sels2016b}. Therefore
	\begin{equation}
	A(\mu)= \sum_i\alpha_1\sigma_y^i +\alpha_2(\sigma_z^i\sigma_y^{i+1} + \sigma_y^i\sigma_z^{i+1}) + \dots,
	\end{equation}
where the first term is simply a generator of rotations along the XZ plane, and would be the exact gauge potential in the absence of the spin-spin coupling $J$. The ``$\dots$" represents higher-order terms. At large values of $h$ we have $\alpha_1\sim \lambda/2h$ and $\alpha_2\sim \lambda J/2h^2$. As such, the magnitude of the rotation is determined by the \textit{energy gap} in the system, as expected. The AGP is translationally invariant: in general a gauge can be chosen such that it obeys all of the symmetries of the Hamiltonian.
	
	The rotated operators and Hamiltonian can be computed efficiently for large systems as detailed in Appendix \ref{appendix:finite_rotations}, because the rotation and operators are all local. The leading terms are as follows
	\begin{align}
	    \tilde H=&h_1\sigma_z^i + h_2\sigma_y^i\sigma_y^{i+1} + h_3\sigma_x^i\sigma_x^{i+1} + h_4\sigma_z^i\sigma_z^{i+1}\nonumber\\
	    +&h_5\sigma_x + h_6\sigma_x^i(\sigma_z^{i+1} +\sigma_z^{i-1}) + h_7\sigma_y^{i-1}\sigma_x^{i}\sigma_y^{i+1} + \dots
	\end{align}
Dots correspond to higher-order contributions from the further non-local terms that go into any quasi-local Hamiltonian. The first three terms are modified from the original Hamiltonian, while the new term $\sigma_z\sigma_z$ appears as a density-density interaction, with a strength $h_4\sim J\lambda^2/h^2$ for $h\gg J,\lambda$.  The first four terms are block-diagonal in the sectors with fixed number of Jordan-Wigner fermions. The further terms break the block structure and are suppressed with increasing size of the variational ansatz.
	
	The last step is to compute the low-energy matrix elements. For the particular values chosen, the ground state is a polarized product state $|\downarrow\downarrow\dots\downarrow\downarrow\rangle$ and the one- and two-particle subspaces have the same span as the one- and two-spin-flipped sectors. Therefore the matrix elements of ${\tilde H}$ are particularly easy to compute. Remarkably, for $h=3$ the one- and two-particle states are not the lowest-energy states due to the hopping bandwidth, as some three-particle states have lower energies. This observation means that the projected manifold does not describe all low-energy states of $H$. Nonetheless, this projection contains the most relevant states with the largest contributions to the correlation functions, and therefore the method still works very well. The importance of few-particle states is also generally observed in truncated spectrum approaches (see below).
	
\begin{figure}[t]
	    \includegraphics[width=0.8\linewidth]{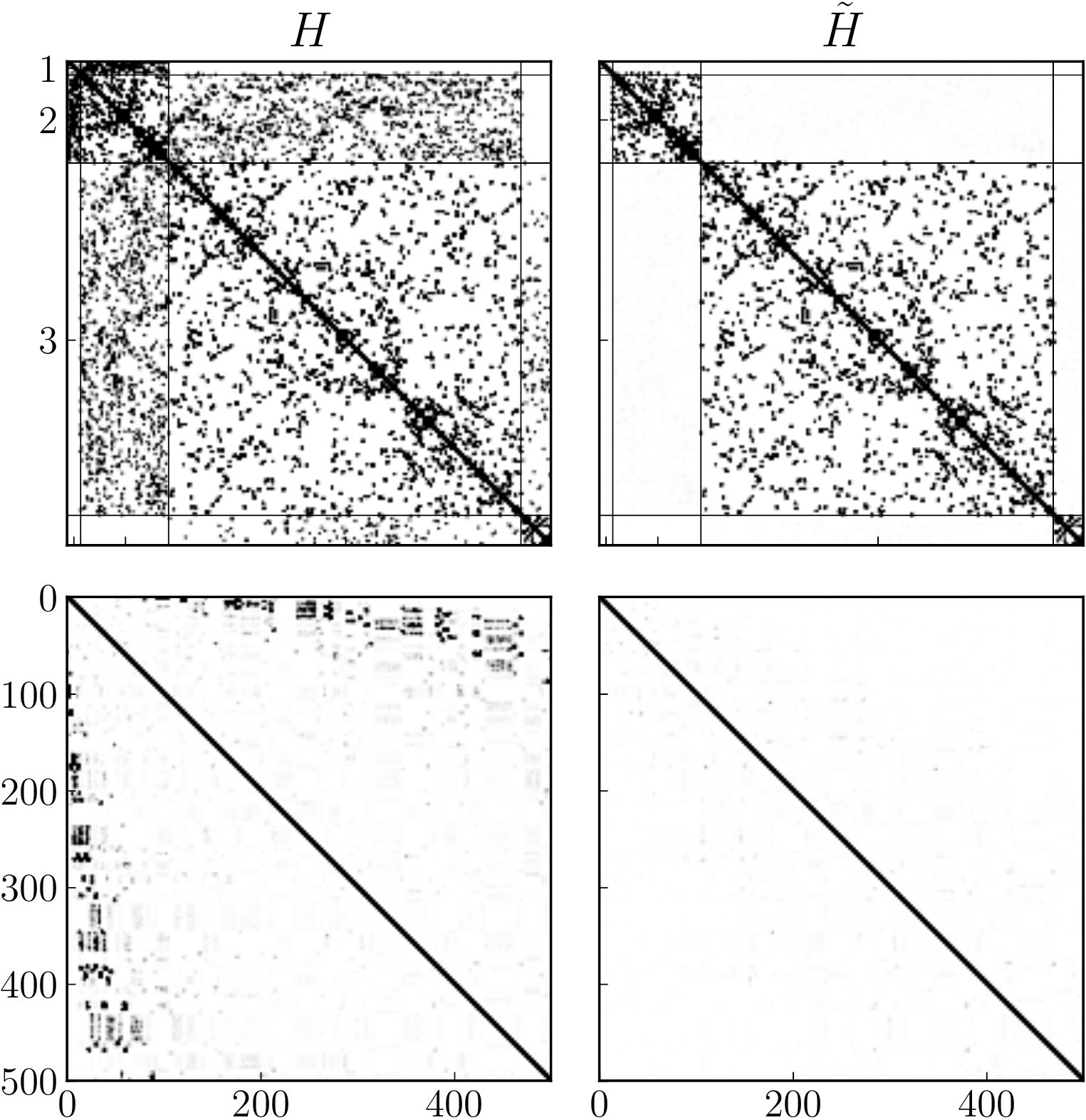}
	    \caption{Illustration of the matrix structure of 14-site original and rotated Hamiltonians in both the eigenbasis of $\sum_i \sigma_z^i$ (top) and the eigenbasis of $H_0$ (bottom).}
	    \label{fig:XY_matrices}
	\end{figure}

    The effective Hamiltonian $\tilde H_\text{eff}$ can be represented in matrix form in the eigenbasis of $H_0$ as
	\begin{align}
	    \begin{bmatrix}
	    \langle \emptyset_0|\tilde H|\emptyset_0\rangle\nonumber &
	    \langle \emptyset_0|\gamma_k\tilde H|\emptyset_0\rangle\nonumber &
	    \langle \emptyset_0|\gamma_k\gamma_k'\tilde H|\emptyset_0\rangle\nonumber\\\\
	    \dots & 
	    \langle \emptyset_0|\gamma_k\tilde H\gamma_{k'}^\dagger|\emptyset_0\rangle\nonumber &
	    \langle \emptyset_0|\gamma_k\gamma_{k'}\tilde H\gamma_{k''}^\dagger|\emptyset_0\rangle\nonumber\\\\
	    \dots & \dots & \langle \emptyset_0|\gamma_k\gamma_{k'}\tilde H\gamma_{k''}^\dagger\gamma_{k'''}^\dagger|\emptyset_0\rangle\nonumber
	    \end{bmatrix}.
	\end{align}
We emphasize that these matrix elements are computed with respect to the unperturbed Hamiltonian's eigenstates, whose analytic properties are known. In principle, computing the overlaps may require a systematic decomposition of $\tilde H$ into products of fermionic raising and lowering operators $\tilde H=\sum\gamma+\gamma\gamma + \gamma\gamma\gamma + \dots$ and repeated application of Wick's theorem; in this example it was avoided by direct computation in the total $Z\in\{-N,-N+1,-N+2\}$ subspace. The effectiveness of the rotation is  shown in Fig.~\ref{fig:XY_matrices}, where the block-diagonal structure can be clearly observed.
	
	It might be tempting to decompose these matrix elements as ``fixed-particle-number" states; however this may only work for low-energy states well separated in energy. At larger energies and at any finite density this particle picture breaks down and states exhibit the chaotic behaviors associated with the Eigenstate Thermalization Hypothesis \cite{rigol2008,DAlessio2015} (also Appendix \ref{app:wignerdyson}).

	Instead, this rotation-and-truncation procedure should be seen as a low-energy/low-density field theory limit of a quantum chaotic model, where integrable states transform into other integrable states. The rotation $U$ can be seen as a dressing of low-energy particle excitations, which include the non-perturbative integrability-breaking effects. For low energies and large gaps, the particle nature of excitations persists, simply from scattering phase-space considerations: a single particle cannot decay into two due to mass differences.
	
\begin{figure*}[t]
	    \centering
	    \includegraphics{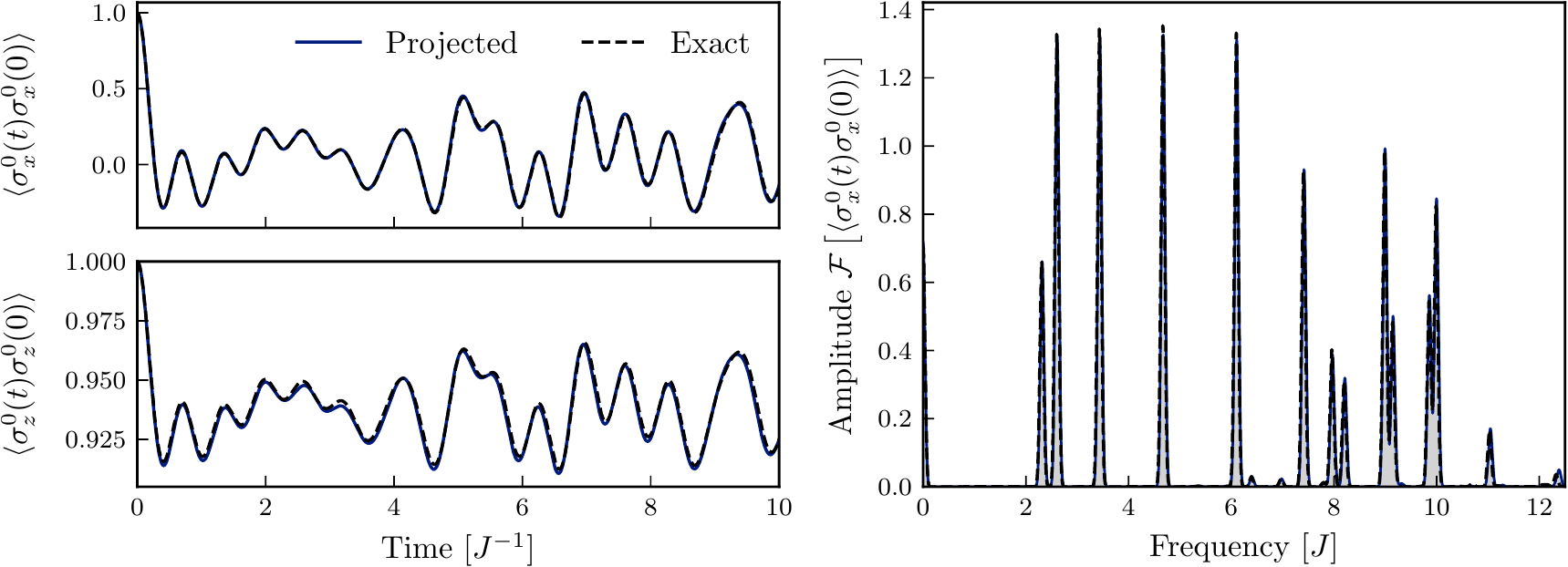}
	    \caption{Approximate and exact response functions for the $U(1)$-broken XY model for a system of 16 sites. Left figures present time-dependent expectation values for $\langle \sigma_\alpha^0(t)\sigma_\alpha^0(0)\rangle$ with $\alpha=x,z$, both for the exact ($2^{16}$ degrees of freedom) and the projected (137 degrees of freedom) system. Observe that exact and projected results are almost indistinguishable. Right figure presents the dynamic structure factor $S(\omega)$ as Fourier transform of $\langle \sigma_x^0(t)\sigma_x^0(0)\rangle$. The rotated Hamiltonian gets both the correct energy eigenvalues and wave function overlaps. A decoherence width of $0.05 J$ has been applied to smoothen the spectrum.}
	    \label{fig:XY_small}
	\end{figure*}

\begin{figure}[h]
	    \includegraphics[width=\linewidth]{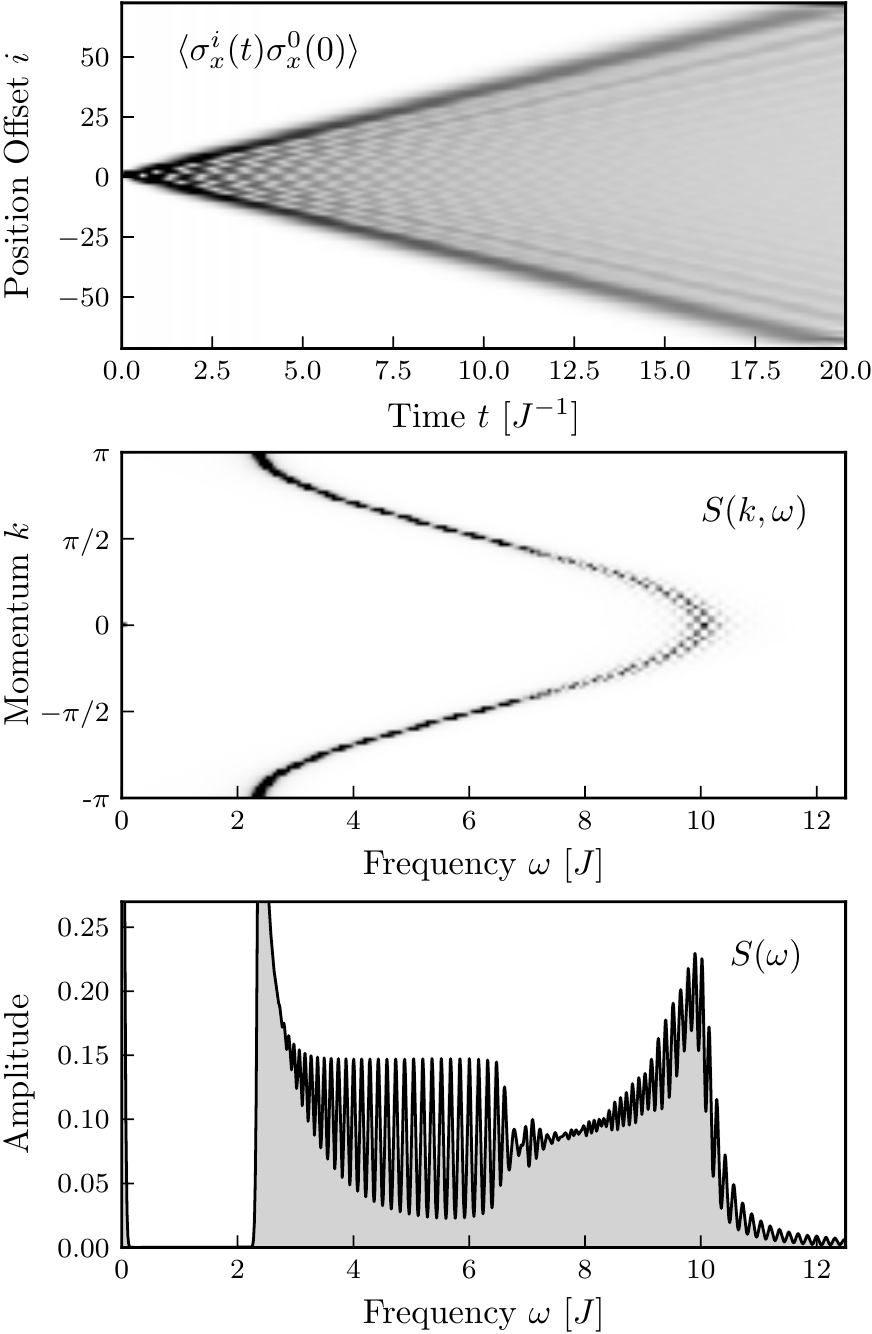}
	    \caption{Offset time correlation function  $\langle \sigma_x^i(t)\sigma_x^0(0) \rangle$ for a system of 144 spin sites and 10441 reduced degrees of freedom. The panels from top to bottom represent (i) the correlation function $\langle \sigma_x^i(t)\sigma_x^0(0) \rangle$, (ii) its spatial- and time-Fourier transform returning the dynamic structure factor $S(k,\omega)$, and (iii) the integrated frequency response $S(\omega)$. Quasi-particle excitations can be clearly observed in all the figures. At low energies, the response is that of a free particle; at larger energies the response widens, signaling finite particle lifetime. In the bottom panel a decoherence factor with width of $0.05J$ has again been applied to smoothen the function.\label{fig:XY_large}}
	\end{figure}

	Results for this model are shown in Figs.~\ref{fig:XY_small} and \ref{fig:XY_large}. The first of these figures shows the comparison between the approximate and exact results for a small system of 16 sites. For computing the exact results the full $2^{16}$ Hilbert space was used, as there are no simultaneous symmetries of the Hamiltonian and wave function. However, the wave function is still close to the ground state and the largest overlap was with low-energy eigenstates, as can be seen in the right figure. The rotated, projected version has $137$ states corresponding to zero, one, or two spins flipped (see also Fig.~\ref{fig:XY_matrices}). Note again that these $137$ states represent the most relevant states, which do not necessarily correspond to the lowest energy states.

	It may come as a surprise how close to the exact result this method is. There is no small parameter in the Hamiltonian, and there are no clearly defined energy spacings. However, one can see even a single-site ansatz can do quite well. Suppose a SW generator with $\theta = \tan^{-1}(\lambda/h)$
	\begin{equation}
	    A = \theta \sum_i\sigma_y^i.
	\end{equation}
	The rotated Hamiltonian can be calculated exactly, as the generator is explicitly local, and is
	\begin{align}
	    \tilde H =\sum_i&\cos^2(\theta)\sigma_x^i\sigma_x^{i+1} +\sin^2(\theta)\sigma_z^i\sigma_z^{i+1} \nonumber\\
	    &+\sin(\theta)\cos(\theta)\big(\sigma_x^i\sigma_z^{i+1} + \sigma_z^i\sigma_x^{i+1}\big) \nonumber\\
	    &+ \sigma_y^i\sigma_y^{i+1} + \sqrt{h^2+\lambda^2}\sigma_z^i.
	\end{align}
	This rotated Hamiltonian still has matrix elements between particle sectors. In particular it contains an anisotropy of $(1-\cos^2\theta)$ corresponding to the difference between the XX and YY interactions, which adds/removes two particles. In addition, the new $XZ$ interactions proportional to $\sin\theta\cos\theta$ allow for the creation of single-particle excitations. However, for $J\ll \lambda,h$ these terms are suppressed by powers of $J/h$ and $J/\lambda$, so the Hamiltonian is still effectively free with a renormalized mass. Higher orders of the gauge potential conspire to make these particle-nonconserving terms smaller, at the expense of adding longer range particle-conserving interactions.

To emphasize applicability of the method to large systems Fig.~\ref{fig:XY_large} presents the response functions for 144 sites, leading to 10441 states in the restricted Hilbert space. At low energies the excitations resemble those of free particles, while at energies greater than $\sim 6 J$, the spectrum broadens as two-particle effects become relevant, indicative of a finite particle lifetime.
	
	This method and these results closely resemble those of \textit{Truncated Spectrum Approximation} (TSA) methods~\cite{Yurov1966,Coser2014,James2018a}, especially when considering the actual procedure for computing the dynamical structure factors. TSA methods aim to provide a description of a non-integrable Hamiltonian by identifying a nearby integrable Hamiltonian and constructing the matrix elements of the original Hamiltonian in a restricted eigenbasis of the integrable one. Here, it is crucial that matrix elements of local operators can be explicitly calculated within such integrable eigenbasis, similar to the SW procedure. The full Hamiltonian is then projected into this subspace, and then numerically diagonalized to get the eigen-spectrum.
	
	The methods presented here are the same, except that the projective subspace is first \textit{rotated} by the Schrieffer-Wolff generator
	\begin{equation}\label{eq:RotatedSubspace_define}
	    U\mathcal PU^\dagger = \tilde{\mathcal{P}}.
	\end{equation}
	Here $\mathcal{P}$ is the projector on $P$, the subspace designated by the symmetries of the integrable Hamiltonian, which would be the basis used for the usual TSA. By associating the generator with the variational gauge potential, the states within $\tilde{P}$ more closely follow the eigenstates of the physical Hamiltonian $H$ by mixing with degrees of freedom outside of the designated symmetry sectors, which would be otherwise inaccessible to the subspace $P$. This is because the generator $A$ does not necessarily share the symmetries of $H_0$, only the symmetries of $H$.

	An additional insight can be drawn from the equivalence of Heisenberg and Schr\"odinger pictures. Under the Heisenberg picture used extensively in this work, operators are rotated into the ``$\sim$" basis (c. f. Eq~\eqref{eq:Htilde_definition}) while the projective subspace remains the same. Free particle states in $P$ evolve under new Hamiltonian which includes effective interaction terms as well as renormalized kinetic energies and masses. Potentially, this rotated Hamiltonian might even be recognized as an integrable field theory in the low-energy limit \cite{DeLuca2016}.
	
	Alternatively, under the Schr\"odinger picture, the Hamiltonian stays the same, while the projective subspace changes (e.g. Eq.~\eqref{eq:RotatedSubspace_define}). This is reminiscent of classical KAM theory \cite{Wayne1996}, where, under an integrability-breaking change of a (classical) Hamiltonian, the integrals of motion are deformed by some canonical transformation on the phase space, retaining the integrable nature of the system. Quantum analogues of KAM \cite{Brandino2015} are an interesting subject of study which go well beyond the scope of this paper. However, we comment on the similarities, where the rotation $U$ might be associated with the equivalent canonical classical transformation, and the subspace $P$ are states which deform slowly, or are otherwise islands of ``non-chaotic" behavior in an otherwise chaotic system. For example, here there are low-energy states, which are not necessarily expected to obey the Eigenstate Thermalization Hypothesis, even if high-energy states do.

	\section{Conclusion}
	Schrieffer-Wolff transformations present a tool for describing effective dynamics of the relevant (for example, low-energy) degrees of freedom of interacting systems. Following a unitary transformation decoupling a low-energy subspace from the rest of the Hilbert space, an effective Hamiltonian can be obtained by projecting this transformed Hamiltonian on the selected subspace.
	
	In this work, we considered Schrieffer-Wolff transformations where the generator was variationally obtained, with the explicit purpose of using the resulting effective Hamiltonian for approximate quantum dynamics. It was argued how both the standard Schrieffer-Wolff transformation and the Wegner flow can be obtained as distinct first-order approximations to the variational approach, and the resulting dynamics were illustrated in two examples. 
	
	First, the disordered Fermi-Hubbard model was considered. The variational generator was approximated using a commutator expansion, and the subspace was that of a fixed number of singly-occupied sites. Here, it was shown that the variational approximation allows for accurate results beyond the reach of standard (perturbative) Schrieffer-Wolff methods. Second, a non-integrable XY spin chain was considered, where the variational generator was constructed out of local operators with a given spatial support. After the initial transformation, the resulting Hamiltonian can be seen as a perturbed integrable one, and the projected subspace was chosen to consist of the eigenstates of this integrable Hamiltonian with a fixed number of particles. This was then shown to be able to return accurate response functions for system sizes beyond the reach of traditional methods.

	\section*{acknowledgments}
	P.W.C. gratefully acknowledges a Francqui Foundation Fellowship from the Belgian American Educational Foundation and support from Boston University's Condensed Matter Theory Visitors program. A.P. and J.W. were supported by NSF DMR-1813499 and AFOSR FA9550-16-1-0334.

	\appendix
	
	\section{Perturbative Schrieffer-Wolff transformation for the Fermi-Hubbard model}
	\label{appendix:SWforFH}
	
	In order to be self-contained, we here explicitly show how the Schrieffer-Wolff transformation of the non-disordered Fermi-Hubbard model in the large-$\Omega$ limit gives rise to the Heisenberg model. The Fermi-Hubbard Hamiltonian is given by
	\begin{equation}
	    H = \Omega \sum_i n_{i, \uparrow} n_{i, \downarrow} -\lambda \sum_{\langle i,j\rangle } c_{j, \sigma}^{\dagger} c_{i,\sigma},
	\end{equation}
	in which $n_{i,\sigma} = c_{i,\sigma}^{\dagger}c_{i,\sigma}$. We consider the strongly-correlated regime $\Omega \gg \lambda$, with the initial subspace being the space spanned by the highly-degenerate eigenstates of $H_0 = \Omega \sum_i n_{i, \uparrow} n_{i, \downarrow}$ with a given eigenvalue (while this eigenvalue is commonly taken to be $0$, the specific eigenvalue will not influence the SW generator, only the projector). 
	
	The goal is to find a rotation $U=e^{-iS}$ such that the matrix elements between eigenstates of $H_0$ with different eigenvalues are suppressed to $\mathcal{O}(\lambda^2)$ in the rotated Hamiltonian $e^{iS}(H_0+\lambda V)e^{-iS}$. For small $\lambda$, the generator will similarly be of order $\mathcal{O}(\lambda)$, and this equation can be linearized to read
	\begin{equation}
	    [V+i[S,H_\Omega],H_\Omega] = 0.
	\end{equation}
	with the kinetic term given by
	\begin{equation}
	   V = -\lambda \sum_{\langle ij \rangle } c_{j, \sigma}^{\dagger} c_{i,\sigma}.
	\end{equation}
	The solution to this equation can be found by rewriting $V$ as
	\begin{equation}
	    V = \lambda\sum_{\langle ij \rangle, \sigma} g_{ij,\sigma} + \left(h^{\dagger}_{ij,\sigma}+h_{ij,\sigma}\right)
	\end{equation}
	with correlated hopping operators
	\begin{align}
	    g_{ij,\sigma} &= n_{j,\overline{\sigma}}c_{j, \sigma}^{\dagger} c_{i,\sigma}n_{i,\overline{\sigma}} \nonumber \\
	    &\qquad + (1-n_{j,\overline{\sigma}})c_{j, \sigma}^{\dagger} c_{i,\sigma}(1-n_{i,\overline{\sigma}}) \\   
	    h^{\dagger}_{ij,\sigma} &= n_{j,\overline{\sigma}} c_{j, \sigma}^{\dagger} c_{i,\sigma} (1-n_{i,\overline{\sigma}}).
	\end{align}
	The first term describes the hopping of doublons and holons -- the eventual quasi-particles within the effective Hamiltonian -- leaving the eigenvalue of $H_0$ invariant, whereas the second term describes the creation (and annihilation) of doublon-holon pairs, changing the eigenvalue of $H_0$ by $\pm \Omega$. This is in analogy with the first example with Hamiltonian $H = H_0 + \lambda(V_+ + V_-)$. Since these operators can only change the eigenvalue by $\pm \Omega$, the SW generator can be found as
	\begin{equation}
	    S = \frac{i}{\Omega^2} \left[H_0,V\right ] = \frac{i}{\Omega}\sum_{\langle ij\rangle,\sigma}\left(h_{ij,\sigma}^{\dagger}-h_{ij,\sigma}\right),
	\end{equation}
	leading to a rotated Hamiltonian (up to $\mathcal{O}(\lambda^3)$)
	\begin{align}
	    e^{iS} \left(H_0 + V\right)e^{-iS} &= H_0 + \lambda \sum_{\langle ij \rangle,\sigma}  n_{j,\overline{\sigma}}c_{j, \sigma}^{\dagger} c_{i,\sigma}n_{i,\overline{\sigma}} \nonumber \\
	    &+\frac{\lambda^2}{\Omega} \sum_{\langle ij\rangle, \sigma}\sum_{\langle kl\rangle, \sigma'}  [h^{\dagger}_{ij,\sigma},h_{kl,\sigma'}].
	\end{align}
	Evaluating the commutator and projecting onto an eigenspace of $H_0$ leads to an effective Hamiltonian (up to an unimportant constant)
	\begin{equation}
	    \tilde{H}_{\text{eff}} = -\lambda \sum_{\langle ij \rangle,\sigma} g_{ij,\sigma}+\frac{4\lambda^2}{\Omega} \sum_{\langle ij \rangle}\left(\vec{S}_i \cdot \vec{S}_j - \frac{n_i n_j}{4}\right),
	\end{equation}
	with spin operators 
	\begin{align}
	    S_i^x &= \frac{1}{2}\left(c^{\dagger}_{i,\uparrow}c_{i,\downarrow}+c^{\dagger}_{i,\downarrow}c_{i,\uparrow}\right), \\
	    S_i^y &= \frac{i}{2}\left(c^{\dagger}_{i,\uparrow}c_{i,\downarrow}-c^{\dagger}_{i,\downarrow}c_{i,\uparrow}\right), \\
	    S_i^z &= \frac{1}{2}\left(c^{\dagger}_{i,\uparrow}c_{i,\uparrow}-c^{\dagger}_{i,\downarrow}c_{i,\downarrow}\right).
	\end{align}
	Away from half-filling, the effective Hamiltonian is given by the $t-J$ Hamiltonian \cite{Bares1994}, while at half-filling the projection of $g_{ij,\sigma}$ vanishes and the effective Hamiltonian simplifies to the Heisenberg Hamiltonian.
	
	\section{Variational minimization}\label{appendix:variational_minimization}
	The variational minimization required to find the generator $A(\mu)$ (c.f. Eq.~\eqref{eq:min_norm}) can be done in two steps. First, make a user-informed guess to some set of operators $\{ B_\alpha\}$. For example, one could choose all operators with support within 3 local spins. The generator is thus given in terms of variational parameters $\{a_\alpha\}$:
	
	\begin{equation}
	A(\mu,\{a\}) = \sum_{\alpha} a_\alpha  B_\alpha.
	\end{equation}
	
	The variational minimization is thus to find the best approximate solution to Eq.~\eqref{eq:Adef2} as
	
	\begin{equation}\label{eq:app:tracemin}
	\text{MIN:}\quad \tr{\big(\big[H,V + i[A(\{a\}),H]\big]\big)^2}.
	\end{equation}

	Plugging in the ansatz above, we find (up to an unimportant constant),
	
	\begin{eqnarray}
	&&-a_\alpha a_\beta \tr{\big[H,[B_\alpha,H]\big]\big[H,[B_\beta,H]\big]}\\
	&&-2ia_\alpha\tr{\big[H,V\big] \big[H,[B_\alpha,H]\big]}\nonumber\\
	&&=a_\alpha M_{\alpha\beta}a_\beta + a_\alpha X_\alpha\nonumber\\
	&&\to\text{MIN:}\quad \vec{\nabla}(a_\alpha M_{\alpha\beta}a_\beta + a_\alpha X_\alpha)=0\nonumber\\
	&&a_\alpha = \big(M_{\alpha\beta}+M_{\alpha\beta}^T\big)^{-1}X_\beta
	\end{eqnarray}
	
	Thus, the minimization is equivalent to finding the inverse of a matrix of rank of the ansatz size.
		
	\section{Generating the transformed Hamiltonian}\label{appendix:finite_rotations}
	
	One of the numerical challenges of this work was computing the rotated Hamiltonian $\tilde H$ and associated wavefunctions and observables. The evolution of the operators is written as
	\begin{equation}
	    \partial_\mu Q(\mu) = i[Q(\mu),A(\mu)]
	\end{equation}
	The challenge is, of course, in implementing this evolution. There are several ways.

	\begin{itemize}
	    \item \textbf{Matrix Product Operators}. Because the rotated operator is quasi-local, one may do time-evolution of operators under a matrix product operator ansatz with a reasonably small bond dimension.
	    
	    \item \textbf{Krylov Subspaces}. Using super-operator formalism one may evolve operators in a subspace, which amounts to a low-order resummation of a Baker-Campbell-Hausdorff expansion \cite{Viswanath2008}.
	    
	    \item \textbf{Exact Evolution on small systems}. If the evolution length is small enough, an exact evolution can be done for a smaller system, then expanded. This is the method used in this work. As such, more details are given below.
	    
	\end{itemize}
	
	The translationally-invariant operator $\tilde H$ is computed in the following way. First, pick the translationally-invariant terms (say, the hopping $\sigma_x^i\sigma_x^{i+1}$) and act with them on two center spins of (nominally) 12 spins as some dense $2^{12}\times 2^{12}$ operator. Then, implement time evolution of Eq.~\eqref{eq:Q_evolution} to find the $2^{12}\times 2^{12}$ dense operator representing $\tilde H$. Next, compute the decomposition of $\tilde H$ into Pauli matrices via traces, using the identity
	
    \begin{equation}
        \tilde H = \frac{1}{2^N}\sum_{\{\alpha\}}\tr{\sigma_\alpha^i\tilde H}\sigma_\alpha^i + \tr{\sigma_\alpha^i\sigma_\beta^j\tilde H}\sigma_\alpha^i\sigma_\beta^j + \dots
    \end{equation}
Here the summation runs over all of the $4^N$ combinations of the Pauli matrices, which form a complete, trace-orthogonal operator basis. Because the operator $\tilde H$ is known to be quasi-local, one would expect most of the contributing Pauli operator strings to only have a few operators: because of this, using a sub-basis of all Pauli operator strings of extent less then, say, 5 sites captures $\tilde H$ almost exactly.
	
With this set of translationally-invariant operators in hand, it is simple to replicate across the larger system: the $m$ operators of the rotated translationally invariant term becomes $mN$ operators across $N$ sites. Similar challenges exists in computing the rotated wave function $\ket{\tilde \psi}=\mathcal P U^\dagger|\psi\rangle$, as one must rotate \textit{then} project. For this work, evolution was done via a Krylov subspace on sparse vectors.
	
	\begin{figure}[t]
	    \includegraphics[width=\linewidth]{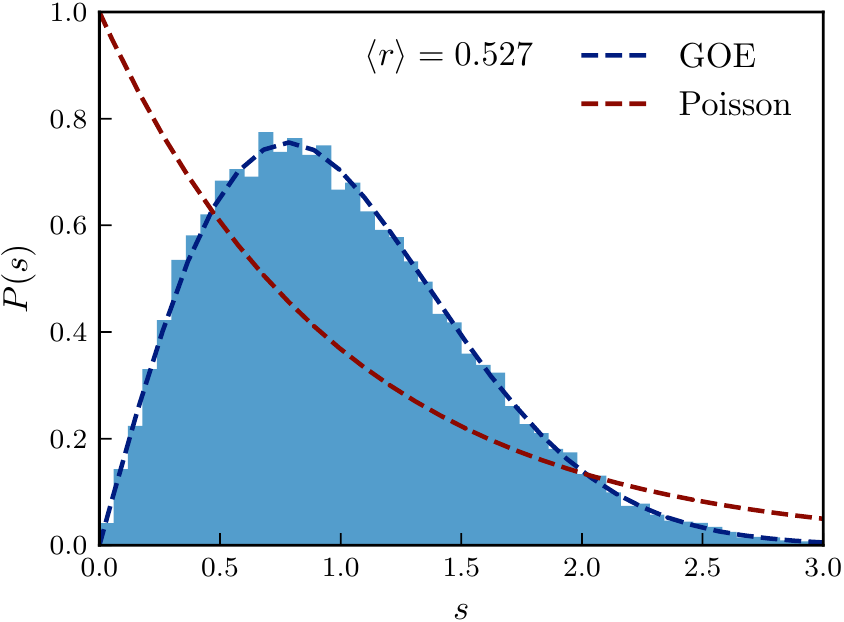}
	    \caption{Level spacing statistics $P(s)$ with $s_n = E_{n+1}- E_n$ for the zero-momentum sector of a 20-spin XY model in a longitudinal field (see text for details). The blue dashed line represents the Wigner-Dyson statistics characteristic of chaotic models, whereas the red line represents the Poisson distribution characterizing integrability. The average value of $r_n= \text{min}(s_n,s_{n+1})/ \text{max}(s_n,s_{n+1})$ also returns a value close to the expected Wigner-Dyson value of  $\braket{r}=0.536$ \cite{Guhr1998}. }
	    \label{fig:XY_WignerDyson}
	\end{figure}
	
	\section{U(1) Broken XY Model is Quantum Chaotic}
	\label{app:wignerdyson}

	It is simple to check that we are not missing out on additional symmetries of the $U(1)$-symmetry-broken XY model and that it is quantum chaotic, in that the spectral statistics follow a Wigner-Dyson distribution in the middle of the spectrum \cite{Guhr1998,DAlessio2015}. The only two symmetries are parity and translation, and so one can compute the statistics within a given symmetry sector. In Fig.~\ref{fig:XY_WignerDyson} is the level spacing statistics presented for the zero-momentum, parity +1 sector of Hamiltonian Eq.~\eqref{eq:U1_XYHamiltonian}, with $J_{xx}=J_{yy}=1$, $h=3$, $\lambda=1.25$, and 20 sites. These statistics and the correponsnce to the Wigner-Dyson distribution indicate that this model is in fact quantum chaotic.

	\bibliographystyle{apsrev4-1}
	\bibliography{citationlist}

\end{document}